\documentclass[fleqn,10pt]{wlscirep}
\usepackage[utf8]{inputenc}
\usepackage[T1]{fontenc}
\usepackage{makecell}

\usepackage{soul}

\title{Compact Spectroscopy of keV to MeV X-rays from a Laser Wakefield Accelerator}
\author[1]{A. Hannasch}
\author[2]{A. Laso Garcia}
\author[1,2]{M. LaBerge}
\author[1]{R. Zgadzaj}
\author[2]{A. K\"{o}hler}
\author[2]{J. P. Couperus Cabada\u{g}}
\author[2]{O. Zarini}
\author[2]{T. Kurz}
\author[2]{A. Ferrari}
\author[2]{M. Molodtsova}
\author[2]{L. Naumann}
\author[2]{T. E. Cowan}
\author[2]{U. Schramm}
\author[2]{A. Irman}
\author[1,*]{M. C. Downer}

\affil[1]{The University of Texas at Austin, Department of Physics, Austin, Texas 78712-1081, USA.}
\affil[2]{The Helmholtz-Zentrum Dresden-Rossendorf, Institute for Radiation Physics, 01328 Dresden, Germany}
\affil[*]{downer@physics.utexas.edu}

\begin{abstract}
We reconstruct spectra of secondary X-rays from a tunable 250-350 MeV laser wakefield electron accelerator from single-shot X-ray depth-energy measurements in a compact (7.5 $\times$ 7.5 $\times$ 15 cm), modular X-ray calorimeter made of alternating layers of absorbing materials and imaging plates.  X-rays range from few-keV betatron to few-MeV inverse Compton to >100 MeV bremsstrahlung emission, and are characterized both individually and in mixtures.  Geant4 simulations of energy deposition of single-energy X-rays in the stack generate an energy-vs-depth response matrix for a given stack configuration.  An iterative reconstruction algorithm based on analytic models of betatron, inverse Compton and bremsstrahlung photon energy distributions then unfolds X-ray spectra, typically within a minute.  We discuss uncertainties, limitations and extensions of both measurement and reconstruction methods. 
\end{abstract}
\begin{document}

\flushbottom
\maketitle
%
%
\thispagestyle{empty}

\noindent
Accelerator-based sources of bright, hard X-rays have enabled decades of advances in materials science \cite{Jaeschke2016SynchrotronLasers}, medicine \cite{Lewis1997MedicalX-rays,Suortti2003MedicalRadiation}, geology \cite{Ketcham2001AcquisitionGeosciences}, warm dense matter science \cite{Falk2014CombinedShock-and-release}, radiography of high-$Z$ materials \cite{Chen2007Dual-energyDetection} and non-destructive testing in industry \cite{Hanke2008X-rayCharacterization}. The radio-frequency electron accelerators that underlie these sources, however, are limited to accelerating gradients of $\sim 100$ MeV/m \cite{Allen1989High-GradientKlystron}. Consequently they are tens to hundreds of meters long, expensive to build and operate and challenging to access. Laser wakefield accelerators (LWFAs) powered by intense laser pulses interacting with a plasma \cite{Tajima1979LaserAccelerator,Esarey2009PhysicsAccelerators} offer tabletop complements to conventional accelerators, but require a unique set of diagnostics \cite{Downer2018DiagnosticsAccelerators}. With accelerating gradients of $\sim 100$ GeV/m, LWFAs can accelerate electron bunches within several cm to energies $E_e$ approaching $10$ GeV \cite{Gonsalves2019PetawattWaveguide}, with bandwidth $\Delta E_e/E_e \sim \,$1-15\% and charge $Q \sim 100$s of pC.  LWFAs are emerging as versatile small-laboratory sources of fs hard X-ray pulses \cite{Corde2013FemtosecondAccelerators}, with photon energies and peak brilliance rivaling those of their conventional synchrotron counterparts, and with a growing list of applications \cite{Albert2014LaserRequirements,Kneip2011X-rayAccelerator,Dopp2016AnImaging}. 

LWFAs can generate three types of secondary X-rays \cite{Corde2013FemtosecondAccelerators}: betatron, inverse Compton scattered (ICS) radiation, and bremsstrahlung. Betatron radiation originates from transverse undulations of accelerating electrons in a wake's focusing fields, and is a natural byproduct of the acceleration process  \cite{Esarey2002SynchrotronChannels,Kostyukov2003X-rayChannel}.  A LWFA producing 250-350 MeV electrons emits betatron X-rays with a synchrotron-like spectrum, with critical energy $E_c \sim$ several keV \cite{Rousse2004ProductionInteraction}.  ICS radiation results from backscatter of counter-propagating laser photons of energy $E_L$ from accelerated electrons of Lorentz factor $\gamma_e$, upshifting the photons to energy $E_x \sim 4 \gamma_e^2 E_L$ \cite{Esarey1993NonlinearPlasmas}. Thus ICS of $E_L = 1.5$ eV photons from electron bunches with peak energy in the range $250 < E_x < 350$ MeV ($500 < \gamma_e < 700$) generates X-rays with spectral peaks in the range $0.5 < E_x < 2$ MeV.  Bremsstrahlung X-rays result from collisions, and associated acceleration, of relativistic electrons passing through a converter after the accelerator, producing broadband X-rays with photon energy up to $E_e$ \cite{Glinec2005High-resolutionSource}.  Secondary X-ray photons from LWFAs thus span an energy range from several keV to several hundred MeV, enabling a wide range of applications \cite{Corde2013FemtosecondAccelerators,Albert2014LaserRequirements}, but requiring an unusually versatile spectrometer for source characterization \cite{Albert2014LaserRequirements}. 

Currently multiple types of spectrometers are required to cover the photon energy range of X-rays from LWFAs. For $E_x \leq 20$ keV, X-ray-sensitive charge-coupled devices (CCDs) operating as photon counters can build up a histogram of the spectrum of low-flux X-rays by measuring the charge that individual X-ray photons deposit in single pixels or pixel groups \cite{Fourment2009BroadbandSources, Kohler2016Single-shotAccelerator}. For $1 \leq E_x \leq 90$ keV, Ross filter pair arrays, which take advantage of the wide distribution of K-edge absorption energies across the periodic table, can analyze the spectral content of X-rays in a single shot \cite{Ross1928AX-Radiations,Khutoretsky1995DesignKeV}. For $90 < E_x < 500$ keV, the sharp absorption sensitivity of K-edges is left behind, but broader differential transmission curves of high-$Z$ materials still enable lower-resolution spectral analysis \cite{King2019X-rayLasers}. For $E_x > 1$ MeV, differential transmission detectors lose resolution quickly, and Compton scattering and $e^+ e^-$ pair production become the main processes for resolving X-ray photon energy \cite{Singh2018CompactExperiments,Tiwari2019GradientRange}.  
X-rays of $E_x > 1$ MeV impinge on a converter, generating Compton electrons and/or $e^+ e^-$ pairs that are energy-analyzed in a magnetic spectrometer.  The energy of the secondary Compton electrons and $e^+ e^-$ pairs is related straightforwardly to that of the incident X-rays, provided the converter is thin enough to avoid multiple scattering events.  This converter thickness requirement limits signal-to-noise ratio, often necessitating averaging over multiple shots.  
To date, Compton/pair-production spectrometers have only measured broadband X-ray spectra. They have not yet measured peaked spectra, e.g. from linear ICS \cite{Yan2017High-orderScattering}. 

Here, we spectrally characterize betatron, bremsstrahlung and ICS X-rays from a 250 - 350 MeV LWFA in a single shot, using a single, compact, inexpensive instrument:  a modular calorimeter consisting of a stack of absorbers of varying $Z$ and thickness, interlaced with imaging plates (IPs). The present measurements utilized a single fixed stack design to analyze an unprecedented 4-decade photon-energy range, demonstrating the spectrometer's universality.  However, the design is easily modified to enhance sensitivity and/or resolution within a narrower spectral range of interest.  The current geometry can diagnose energies as low as $\sim7$ keV, typical of betatron radiation, and as high as 100 MeV, typical of thick target bremsstrahlung radiation. We reconstruct spectra that are betatron-, ICS- or bremsstrahlung-dominated, as well as spectra containing a mixture of different types of X-rays with widely separated photon energies. 

%
%

\begin{figure}[!ht]
\centering
\includegraphics[width=.9\textwidth]{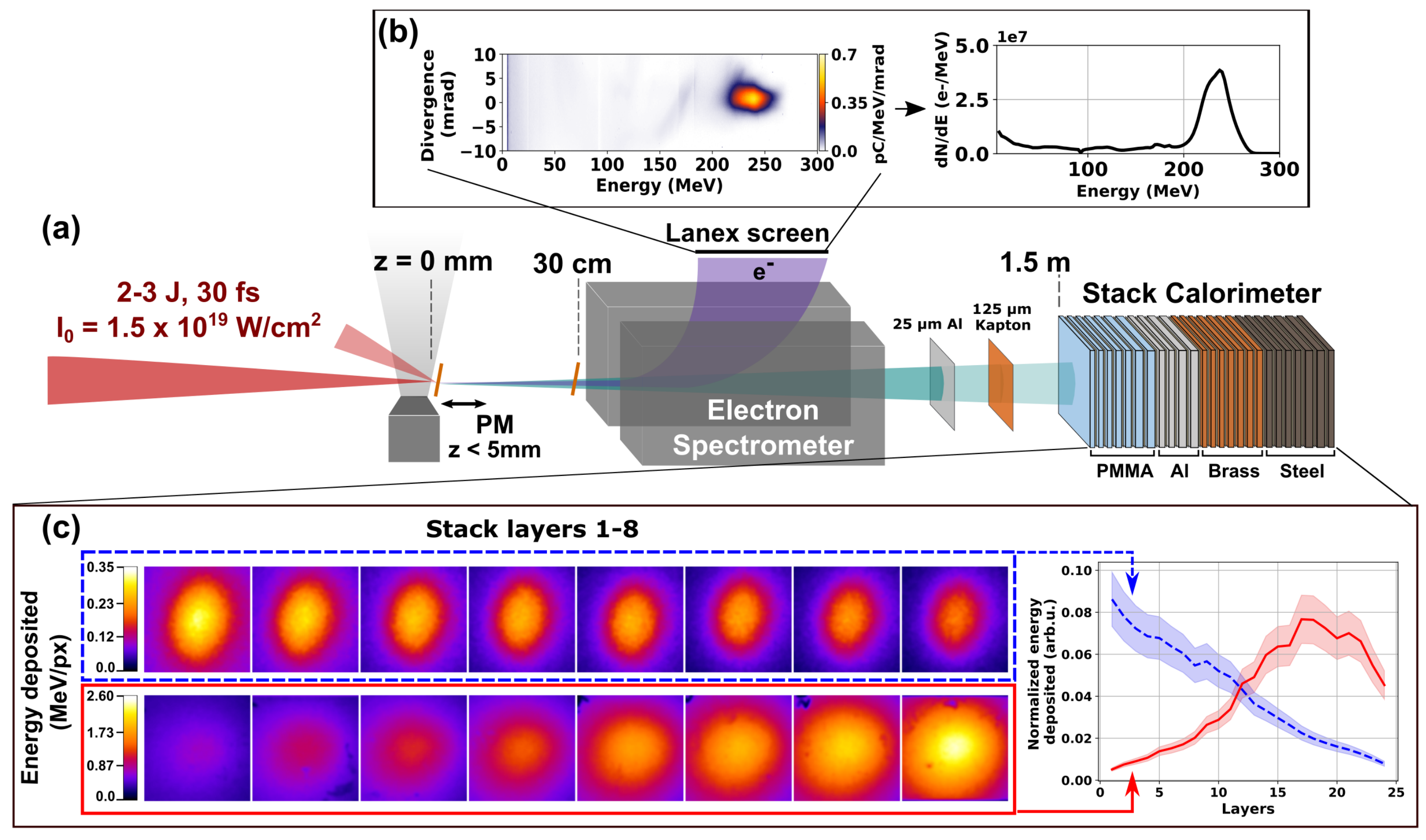}
\caption{\label{fig:experimental schematic} LWFA X-ray spectrometry overview.  (a) Schematic set up showing (left to right) incident laser pulse, gas jet, tilted plasma mirror (PM) positioned at $0 < z < 0.5$\,cm from gas jet exit for generating ICS X-rays, converter at $z \approx 30$\,cm for generating bremsstrahlung, 1 T magnetic electron spectrometer and X-ray stack calorimeter outside vacuum chamber.  (b) Representative single-shot electron spectrum.  Left:  raw data from luminescent Lanex screen.  Right:  electron energy distribution integrated over emission angle. (c) Two depth-energy distributions from calorimeter.  Top left (dashed blue box):  first 8 image plate exposures for ICS-dominated radiation, generated with $25\,\mu$m-thick, low-$Z$ PM at $z = 0.1$ cm.  Bottom left (solid red box):  same for bremsstrahlung-dominated radiation, generated with 800\,$\mu$m-thick, high-$Z$ converter at $z = 30$ cm. Color bars:  relative scaling of deposited energy.  Right:  corresponding color-coded plots of transversely-integrated deposited energy (normalized to total deposited energy) vs.\,layer number for 24 layers.  Shaded regions:  calibration uncertainty (see Supplementary Material)}
\end{figure}

The calorimeter used here is based on a design developed by Garcia \textit{et al.} (2021) for spectrally analyzing various hard X-ray sources, including ps X-ray pulses from intense laser-solid interaction and natural X-ray emission \cite{LasoGarcia2021MultipurposeInteractions}.  X-ray calorimeters consisting of alternating absorbers and detectors were used in some previous laser-plasma experiments \cite{Chen2008ADosimeters,Albert2013AngularAccelerator,Scott2013MeasuringDetectors,Horst2015AInteraction,Rhee2016SpectralPlasma}. A calorimeter stack for diagnosing LWFA electron spectra, charge and divergence was also previously reported \cite{Hidding2007NovelEnvironment}, although magnetic spectrometers are now universally used for this purpose.
The present study \emph{differs} from prior work in the following key respects:  (\textit{i}) It focuses exclusively on secondary X-rays from LWFAs.  (\textit{ii}) It measures and reconstructs a wider range of photon energies, by employing a longer stack (24 layers) and a strategic mixture of low- and high-Z absorbers.  We thereby extend the work of Albert et al. \cite{Albert2013AngularAccelerator} by unfolding not only $\sim 10$ keV betatron X-ray spectra, but also MeV ICS and bremsstrahlung spectra from the same LWFA. (\textit{iii}) It streamlines spectral reconstruction by employing an algorithm based on physical or empirical models of betatron, ICS and bremsstrahlung emission and least squares optimization.  The resulting fast convergence (currently $\ll 1$ minute) on a standard lab-grade laptop could, with improvements, become suitable for analyzing data in real time from rep-rated LWFAs.  It thereby differs from the more general Bayesian reconstruction algorithm \cite{DAgostini1995ATheorem} employed in Garcia \textit{et al.}\,(2021), which can in principle reconstruct and discover unexpected spectral features different in form from the initial guess, but which takes longer to converge.  An additional streamlining element is that we fit spatially-integrated depth-energy distributions, and reconstruct spatially-integrated spectra, foregoing the angular resolution of betatron energies employed in Albert \textit{et al.}\,(2013) \cite{Albert2013AngularAccelerator}. Nevertheless, if necessary the reconstruction algorithm could easily be generalized for applications not requiring rapid analysis.

%
%

\section*{Results}
\textbf{Generation and diagnosis of X-rays.} Fig.\,\ref{fig:experimental schematic}(a) presents a schematic overview of the LWFA X-ray spectrometry setup. A high-energy, ultra short laser pulse impinged on a nitrogen doped helium gas jet and excited a laser wakefield that accelerated electrons (see Methods). A magnetic electron spectrometer dispersed these electrons and diagnosed their energy distribution. A stack calorimeter consisting of 24 absorbing layers interspersed with IPs, recorded the depth-energy distribution of particle cascades initiated by secondary X-rays from the LWFA. Supplementary Table S1 lists absorber compositions and thicknesses and IP parameters for the stack used here.  We generated and characterized four types of X-ray outputs:  

\begin{enumerate}
\item \textit{Pure betatron X-rays}.  Betatron X-rays, generated in a 3-mm jet, propagated from LWFA exit ($z=0$) to calorimeter (entrance plane at $z=150$\,cm), passing only through a $25\,\mu$m-thick Al laser blocking foil and a $125\,\mu$m-thick Kapton vacuum chamber window, both downstream of the $e$-spectrometer, which together blocked $<\,7$\,keV X-rays.  The $e$-beam generated no other X-rays outside the LWFA.  We cross-checked unfolded betatron X-ray spectra in two ways:  (a) by measuring betatron X-ray spectral histograms independently on separate, but similar, shots using a Pixis-XO 400BR photon-counting CCD sensitive to X-ray photon energies up to $\sim30$ keV \cite{Kohler2016Single-shotAccelerator}; (b) by simulating the spectra generated by a single electron with various trial oscillation trajectories $r_\beta(t)$ using the classical radiation code CLARA \cite{Pausch2014HowFrameworks} (see Supplementary Material). 
\item \textit{Pure bremsstrahlung}.  We used a 5-mm jet to maximize electron and photon energy, and inserted a thick, high-$Z$ foil (e.g. 800\,$\mu$m-thick Ta) at $z \sim\,30$\,cm, which acted as a converter.  Electrons entering the foil underwent collisions, generating forward bremsstrahlung.  The foil blocked betatron X-rays completely. 
\item \textit{Bremsstrahlung + betatron X-rays}.  We inserted a thin, low-$Z$ foil (e.g. $25\,\mu$m-thick Kapton) at $z \sim\,30$\,cm.  It generated $\sim120\times$ weaker bremsstrahlung, but transmitted most of the incident betatron X-rays.  Thus the two had comparable flux at the detector.  
\item \textit{ICS X-rays}.  We inserted the thin, low-$Z$ foil at $0 < z < 0.5$\,cm.  Here, the transmitted LWFA drive pulse was intense enough to ionize it, converting its front surface to an overdense plasma, or plasma mirror (PM), that retro-reflected the drive pulse back onto trailing electrons, generating ICS X-rays \cite{TaPhuoc2012All-opticalSource,Tsai2015CompactMirror,Dopp2016AnImaging}.  In this configuration, ICS X-rays dominated over betatron/bremsstrahlung background.  Plasma mirroring, and thus ICS, were negligible for foils at $z = 30$\,cm.   
\end{enumerate}

Fig.\,\ref{fig:experimental schematic}(c) contrasts transverse energy profiles recorded by the first 8 IPs for an ICS-dominated (top left, blue dashed box) and a bremsstrahlung-dominated (bottom left, red solid box) shot. The plot on the right side of Fig.\,\ref{fig:experimental schematic}(c) shows transversely-integrated deposited energy vs. layer number for all 24 layers (see table\,\ref{dose analysis} in Methods for details on integration radius and total integrated energy for each source). These markedly different longitudinal energy profiles provide raw data for distinguishing the energy content of the two X-ray pulses.  
%
%
\begin{figure}[!ht]
\centering
\includegraphics[width=.65\textwidth]{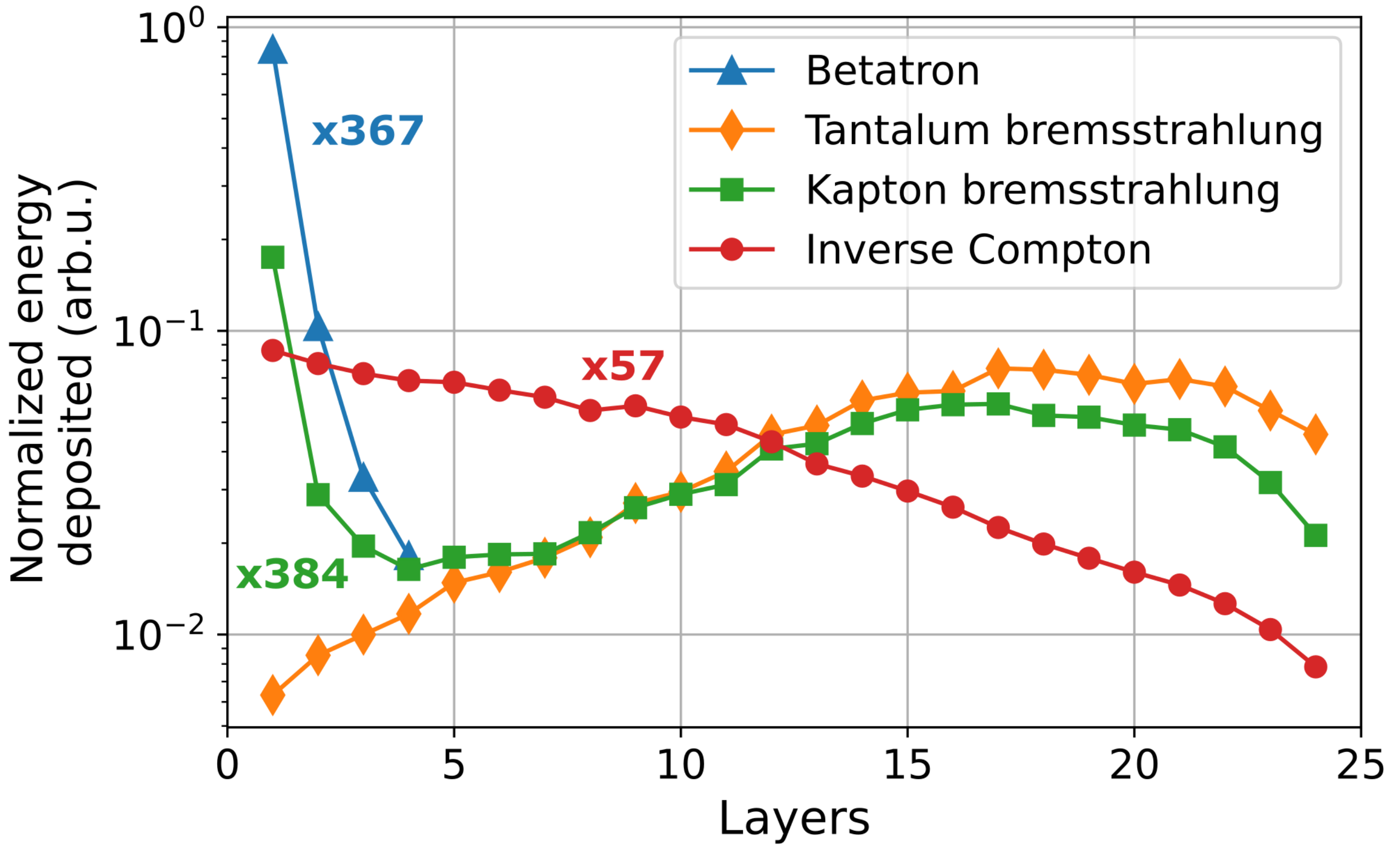}
\caption{\label{fig:Dose comparison} Longitudinal profiles of deposited energy, normalized to total energy deposited in the stack, for each of four LWFA X-ray outputs.  Scaling factors next to each curve indicate that the plotted energy deposition profile was multiplied by the indicated number to give its correct amplitude relative to tantalum bremsstrahlung X-rays (orange diamonds).  
}
\end{figure}

Fig.\,\ref{fig:Dose comparison} compares normalized longitudinal energy profiles for the four X-ray outputs described above. Each data is multiplied by the factor shown to give its true amplitude relative to the pure bremsstrahlung source. Pure betatron X-rays (blue triangles) deposit energy with progressively decreasing amplitude only in the first 4 layers, indicative of the short absorption depth of few-keV photons.  The energy profile of mixed bremsstrahlung/betatron X-rays (green squares, ``Kapton bremsstrahlung'') displays the same sharply-decaying betatron X-ray feature in the first few layers, but now augmented with broadly-distributed deposition deeper in the stack (peaking at layers 16-17) by higher-energy bremsstrahlung photons. Pure bremsstrahlung from a thick, high-$Z$ foil (orange diamonds, ``Tantalum bremsstrahlung'') generates no betatron feature in layers 1-4, only the characteristic broad ``bremsstrahlung'' peak in deeper layers, now stronger by a factor $\sim\,400$.  ICS X-rays (red circles, ``Inverse Compton'') deposit energy in a pattern distinct from the previous cases:  energy deposition decreases monotonically throughout the stack. It is possible to recognize different classes of X-rays immediately from these ``fingerprint'' energy profiles alone, even before analyzing them to reveal their widely differing energy content quantitatively. The multiplicative factors illustrate the high dynamic range of the detector, which shows no saturation over a factor of nearly $500$ in deposited energy. To the best of our knowledge, this is the first direct observation of the three different LWFA X-ray sources and their energy signatures from a single detector.



\vspace{0.2cm}
\noindent
\textbf{Betatron X-rays.} 
The betatron radiation spectrum is derived \cite{Esarey2002SynchrotronChannels} from Li\'{e}nard-Wiechert potentials of accelerating electrons undergoing sinusoidal betatron oscillations of wavenumber $k_\beta = k_p/(2\gamma_e)^{1/2}$ and amplitude $r_\beta$ in the focusing fields of a plasma bubble.  Here, $k_p$ 
is the plasma wavenumber.  When the betatron strength parameter $a_\beta = \gamma_ek_\beta r_\beta$, analogous to a wiggler parameter, exceeds unity (for our experiments, $5 \leq a_\beta \leq 10$) and varies continuously during acceleration, radiation is generated in a forward-directed continuum of overlapping harmonics of the Doppler-upshifted betatron frequency $2\gamma_e^2ck_\beta/(1+a_\beta^2/2)$ up to a critical frequency $\omega_c = 3\gamma_e^3k_\beta^2cr_\beta$, beyond which intensity diminishes. The spectrum of radiation along the axis from a single electron then takes the form \cite{Esarey2002SynchrotronChannels}

\begin{equation}\label{betatron model}
\frac{dN}{d(\hbar\omega)} = C \frac{\omega}{\omega_c^2} K_{2/3}^2\left(\frac{\omega}{\omega_c}\right),
\end{equation}

\vspace{0.2cm}
\noindent
where $C \approx 3 N_\beta e^2 \gamma_e^2 \Delta \Omega/(\hbar^2 \pi^2 \epsilon_0 c)$, $N_\beta$ is the number of betatron periods, $\Delta \Omega$ is an integrated solid angle and $K_{2/3}$ is a modified Bessel function. Here, we constrain the betatron photon spectrum to the form of Eq.\,\ref{betatron model}, and use $\omega_c$ as a fit parameter.

\begin{figure}[!ht]
\centering
\includegraphics[width=.98\textwidth]{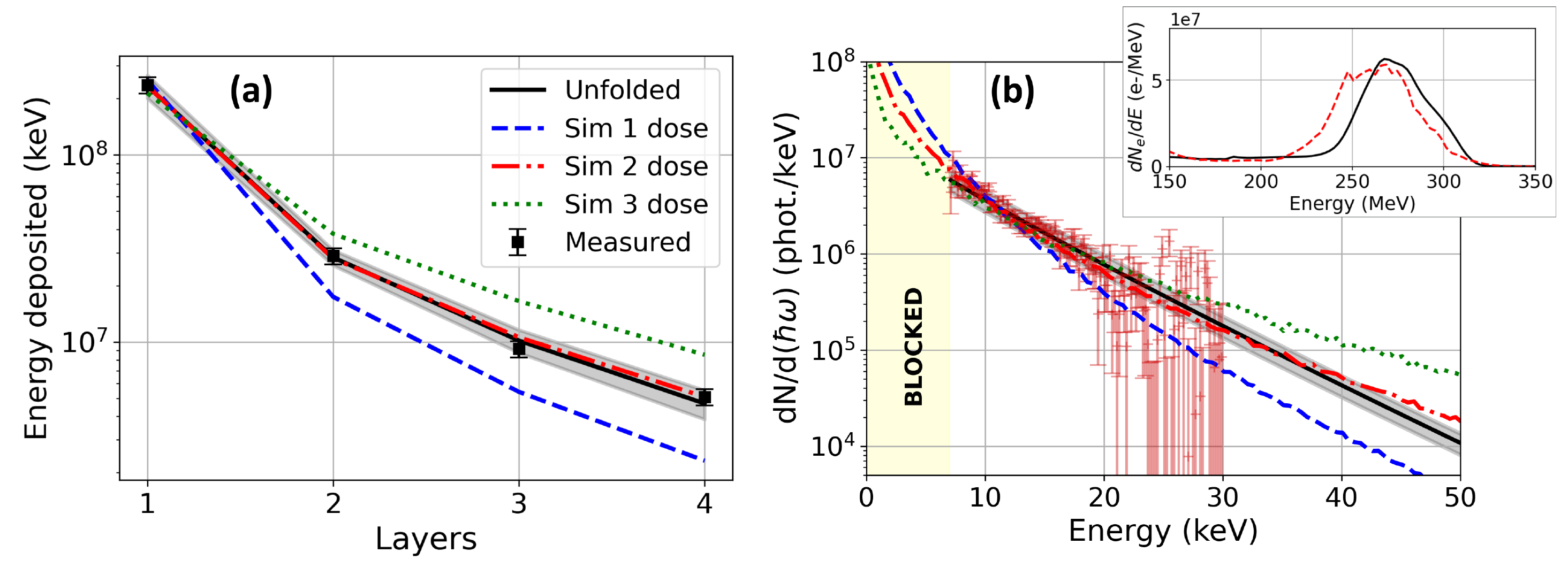}
\caption{\label{fig:betatron only plots}Betatron X-ray results.  (a) Measured (black squares), unfolded (black solid curve) and simulated (colored curves labeled Sim 1, 2, 3) energy deposited in first 4 calorimeter stack layers. (b) Corresponding unfolded spectrum (solid black curve) and uncertainty (grey), compared to betatron spectrum measured independently by X-ray photon counting (red data points).  Colored curves:  CLARA2 simulations of betatron X-ray spectra for $e$-trajectories $r_\beta(t)$ corresponding to final electron energy $E_e$ and oscillation amplitude $r_{\beta 0}$, respectively, of 280\,MeV, $0.5\,\mu$m (blue-dashed, Sim 1); 280\,MeV, $0.9\,\mu$m (red-dashed, Sim 2); 340\,MeV, $0.9\,\mu$m (green-dotted, Sim 3).  Inset:  electron spectra for calorimeter (black) and photon-counting (red) measurements.}
\end{figure}

Data points (squares) in Fig.\,\ref{fig:betatron only plots}(a) show a typical measured on-axis energy deposition profile $D_i^{(meas)}$ ($1 \leq i \leq 4$) from betatron X-rays generated by a 274 MeV ($\gamma_e = 536$) electron bunch with 18 MeV FWHM energy spread [see spectrum in inset of Fig.\,\ref{fig:betatron only plots}(b), black curve] in $n_e = 5\times10^{18}\,\text{cm}^{-3}$ plasma, compared to the unfolded energy distribution $D_i^{(calc)}$ [solid black curve in panel (a)]. We obtain best fit to the measured energy deposition with a X-ray photon spectrum $\frac{dN}{d(\hbar\omega)}(\hbar\omega,\hbar \omega_c)$ of critical photon energy $\hbar\omega_c = 14 \pm 1.5$ keV, shown also by a solid black curve in the main panel of Fig.\,\ref{fig:betatron only plots}(b). The number of photons within the FWHM of the betatron source is $5.5\pm 1.1\times10^{7}$ photons/keV over 7 keV. Yellow shading in Fig.\,\ref{fig:betatron only plots}(b) indicates energies that are blocked by the beam line elements and grey shading gives uncertainties in the unfolded energy profile (a) and spectrum (b), determined via the procedure described in Methods. 
From $E_c$, $n_e$, and $\gamma_e$, we estimate betatron radius $r_\beta = \omega_c/(3\gamma_e^3k_\beta^2c) \approx 10^{23} \,E_c\,{\rm [keV]}/(\gamma_e^2\,n_e\,{\rm [cm^{-3}]}) = 1.0\pm0.1~\mu$m, or $a_\beta = 7 \pm1$.

Red data points (+'s) in Fig.\,\ref{fig:betatron only plots}(b) show results of a typical independent X-ray spectral measurement using the photon-counting CCD, for a shot under the same conditions that yielded an electron bunch of nearly identical energy [Fig.\,\ref{fig:betatron only plots}(b) inset, red dashed curve]. The X-ray spectrum is corrected for the transmission efficiency of the Al laser blocking foil, the Kapton vacuum chamber window [see Fig.\,\ref{fig:experimental schematic}(a)] and an additional filter that attenuated X-ray flux to less than one photon per pixel. The independently measured and unfolded spectra agree within combined uncertainties in the most sensitive range (8-20 keV) of the X-ray CCD.

%
%

The colored curves in Fig.\,\ref{fig:betatron only plots}(b) [blue dashed, red dot-dashed and green dotted curves] show X-ray spectra for three values of $r_\beta$ and $E_e$, selected from simulations for a range of $r_\beta$, $E_e$ values carried out using the classical radiation code CLARA2 \cite{Pausch2014HowFrameworks} (see Supplementary Material). The red dot-dashed curve, which corresponds to $r_\beta=0.9\pm0.1\,\mu$m and $E_e = 280\pm20$\,MeV (Table \ref{beta table}, second row from the bottom), best matches the unfolded and independently measured X-ray spectra over the sensitive range of the CCD detector.  The stated uncertainties in $r_\beta$ and $E_e$ were generated from an ensemble of simulations, and represent variances from the best-fit values. 
Moreover, this simulated spectrum, when input into equation\,\eqref{dose calculation} using the same response matrix $R_{ij}$ used for the unfolding, yielded a calculated deposited energy [Fig.\,\ref{fig:betatron only plots}(a), red dot-dashed curve, ``Sim 2''] nearly indistinguishable from the measured (squares) and unfolded (solid black curve) energy deposition profiles.  This good agreement corroborates the $r_\beta$ value inferred from the unfolding alone.

\begin{table}[ht!]
    \centering
        \begin{tabular}{ |c||c|c|c|c|c| }
        \hline
        & \multicolumn{2}{c|}{Electron parameters} & \multicolumn{3}{c|}{Unfolded Betatron parameters}\\
        \hline
        &$E_{pk}$ (MeV) & $r_\beta$ ($\mu$m) & $E_c$ (keV) & $r_\beta$ ($\mu$m) & $N_{phot}$\\ 
        \hline
        Unfolded& $274 \pm 18$  & - & $14 \pm 1.5$ & $1.0 \pm 0.1$ & $5.5 \pm 1.1 \times 10^{7}$\\
        \hline
        Sim 1   & $280$ & $0.5$ & $9.9$ & $0.66$ & - \\
        \hline
        Sim 2   & $280 \pm 20$  & $0.9 \pm 0.1$ & $14 \pm 2$ & $0.94 \pm 0.1$ & - \\
        \hline
        Sim 3   & $340$ & $0.9$ & $19.4$ & $0.87$ & - \\
        \hline
        \end{tabular}
    \caption{Unfolded parameters for the betatron model based reconstruction (first row) and the simulated and corresponding unfolded parameters for the simulations labeled "Sim 1", "Sim 2" and "Sim 3". The unfolded betatron parameters include the critical energy, betatron radius $r_\beta$ and number of photons with energy $> 7$ keV and within the FWHM of the beam. The simulations only provide the relative shape of the betatron spectra and do not include the photon number.}
    \label{beta table}
\end{table}

The two additional CLARA2 simulation results shown in Fig.\,\ref{fig:betatron only plots}(b) correspond to $r_\beta = 0.5\,\mu$m, $E_e = 280$\,MeV (blue dashed) and $r_\beta = 0.9\,\mu$m, $E_e = 340$\,MeV (green dotted).  Both fall well outside the uncertainty range of the unfolded X-ray spectrum.  Similarly their calculated energy distributions, shown by ``Sim 1'' (blue dashed) and ``Sim 3'' (green dotted), respectively, in Fig.\,\ref{fig:betatron only plots}(a) fall well outside the uncertainty range of the measured energy. When we ran the single-parameter unfolding algorithm on these calculated energy profiles, treated as synthetic data, we found $E_c = 10\pm1$ keV and $r_\beta = 0.66\pm0.2~\mu m$ for ``Sim 1'' and a $E_c=19\pm2$ keV and $r_\beta=0.85\pm0.09$ for ``Sim 3'', consistent with the original CLARA2 input parameters.  These examples illustrate the degree to which the stack-based unfolding method can resolve betatron X-ray parameters associated with different acceleration conditions.

\vspace{0.2cm}
\noindent
\textbf{Bremsstrahlung X-rays.}  Koch and Motz (1959)\cite{KochH.W.Motz1959BremsstrahlungData} have compiled a comprehensive summary of cross-section approximations and experimental data for bremsstrahlung.  Here we model bremsstrahlung spectra 
using either electron scattering cross-sections derived from the Born approximation or the so-called Kramers' law.  The Born differential cross-section for scattering of relativistic electrons to produce bremsstrahlung of photon energy $\hbar\omega$ has the analytic form (neglecting screening effects)\cite{Bethe1934OnElectrons}

\begin{equation}\label{born xsec main}
\left(\frac{d\sigma}{d(\hbar\omega)}\right)_{Born} = \frac{16}{3}  \frac{Z^2 r_e^2 \alpha}{\hbar\omega} \left( 1 - \frac{\hbar\omega}{E_0}+\frac{3\hbar^2\omega^2}{4 E_0^2}\right)\left[\ln\left(\frac{2E_0(E_0-\hbar\omega)}{m_e c^2\hbar\omega}\right)-\frac{1}{2}\right]. 
\end{equation}

\vspace{0.2cm}
\noindent
Here, $Z$ is the charge of the scattering nucleus, $\alpha$ the fine structure constant, $r_e$ the classical electron radius and $E_0$ the initial electron energy. Monoenergetic electrons passing through a thin ($L/L_0<<1$) low $Z$ target (e.g. 25 $\mu$m-thick Kapton) lose negligible energy, so the bremsstrahlung spectrum, i.e. the number of photons per energy bin $dN/d(\hbar\omega)$, has the form of Eq.\,\eqref{born xsec main}. Here, $L$ is target thickness and $L_0$ the radiation length of the target material. Relativistic electrons ($E_0>>137mc^2Z^{-1/3}$) passing through a thick ($L/L_0\sim 1$), high $Z$ target (e.g. 800 $\mu$m-thick Ta), on the other hand, experience energy-dependent alterations to the scattering cross-section because screening of the nucleus by atomic electrons becomes important in this limit, necessitating a correction to Eq.\,\eqref{born xsec main} (see Supplementary Material). We estimate $dN/d(\hbar\omega)$ by integrating the cross-section over target thickness, or equivalently over electron energy loss, assuming that electrons lose energy continuously to radiation at a rate $dE_0/dx = -E_0/L_0$. 

\begin{equation}\label{Born integration}
    \frac{dN}{d(\hbar\omega)} = n \int_{\hbar\omega}^{E_i} \frac{d\sigma}{d(\hbar\omega)} \frac{dE_0}{(-dE_0/dx)} = n L_0 \int_{\hbar\omega}^{E_i} \frac{1}{E_0} \frac{d\sigma}{d(\hbar\omega)} dE_0 
\end{equation}

\vspace{0.2cm}
\noindent
The integration results in a piece-wise function, in which $dN/d(\hbar\omega)$ differs in form for $\hbar\omega$ greater than or less than the final electron energy $E_f$:  

\begin{subequations}\label{born nphot}
\begin{align}
\left(\frac{dN}{d(\hbar\omega)}\right)_{low} = \frac{C}{\hbar\omega} \left( \ln{\frac{E_0}{E_f}}+\hbar\omega \left(\frac{1}{E_0}-\frac{1}{E_f}\right) - \frac{3}{8} (\hbar\omega)^2 \left(\frac{1}{E_0^2}-\frac{1}{E_f^2}\right) \right) \hspace{1.5cm} (\hbar\omega\ < E_f \leq  E_0)  \\
\left(\frac{dN}{d(\hbar\omega)}\right)_{high} = \frac{C}{\hbar\omega} \left( \ln{\frac{E_0}{\hbar\omega}}+\hbar\omega \left(\frac{1}{E_0}-\frac{1}{\hbar\omega}\right) - \frac{3}{8} (\hbar\omega)^2 \left(\frac{1}{E_0^2}-\frac{1}{(\hbar \omega)^2}\right) \right) \hspace{1.5cm} (E_f\leq \hbar\omega\leq E_0)
\end{align}
\end{subequations}

\vspace{0.2cm}
\noindent
where $C = 16 Z^2 r_e^2 \alpha n L_0/3$. The Born approximation model has been used widely to predict or model the properties of bremsstrahlung in experiments \cite{KochH.W.Motz1959BremsstrahlungData}.

%
%
\begin{figure}[!hbt]
\centering
\includegraphics[width=.98\textwidth]{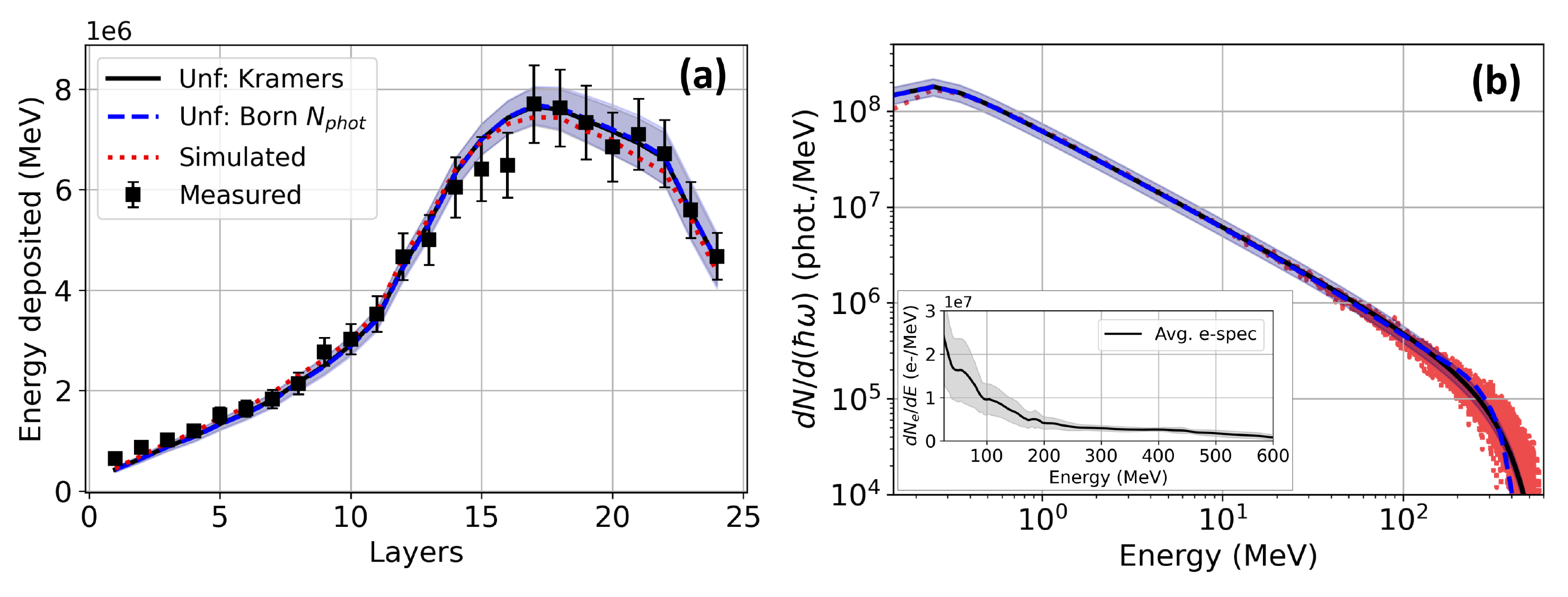}
\caption{\label{fig:brem only plots} Bremsstrahlung X-ray results. (a) Comparison of measured (black squares), simulated (red dotted line) and unfolded energy deposition profiles based on Kramers' law (black solid line) and the Born cross-section (blue dashed line) for the bremsstrahlung dominated case. The corresponding unfolded and simulated spectrum are shown in (b) and the average electron spectrum for the previous 5 shots without the tantalum in the beam path in the inset of (b) with the standard deviation (shaded). The shaded regions in (a) represent the unfolding error and the shaded regions in (b) represent the 20\% uncertainty in the absolute photon number from the calibration.}
\end{figure}

Kramers derived the shape of the bremsstrahlung spectrum by a nonrelativistic semi-classical calculation that considered only continuous electron energy loss, but not discrete electron scattering events or radiation absorption \cite{Kramers1923OnSpectrum}. Nevertheless, a common practice is to approximate the integration of the cross section over energy loss through a thick target using Kramers' law, and to take radiation attenuation within the target into account using NIST attenuation data \cite{NISTmassattenuationdata} 
Moreover, since this integration is equivalent to integrating over incident electron energies, Kramers' model is also widely used to describe bremsstrahlung from thin targets when there is electron energy spread. Kramers' approximation for the bremsstrahlung spectrum has the analytic form

\begin{equation}\label{brems model}
\frac{dN}{d(\hbar\omega)} \approx \frac{C}{\hbar\omega}(E_0 - \hbar\omega)\,, ~~~~~~~~~~ \hbar\omega \leq E_0
\end{equation}

\vspace{0.2cm}
\noindent
where $C=8 \pi^2 N_e r_e Z/(9 \sqrt{3} c \hbar)$ and, for wide electron energy spread, $E_0$ represents the X-ray cutoff photon energy $E_{cutoff}$, which is close to the maximum incident electron energy. In practice, the parameter $E_0$ functions as an empirical parameter for fitting or unfolding spectra.  Kramers' Law has widely and successfully approximated observed bremsstrahlung spectra,
even (its original assumptions notwithstanding) those generated by relativistic electrons in both thick and thin targets  \cite{KochH.W.Motz1959BremsstrahlungData}.

Data points (black squares) in Fig.\,\ref{fig:brem only plots}(a) show a typical energy deposition profile $D_i^{(meas)}$ $(1 \leq i \leq 24)$, integrated over the beam FWHM of 11.5 $\pm$ 0.4 mrad, from bremsstrahlung X-rays that LWFA electrons generated in an 800 $\mu$m-thick Ta target. The inset of Fig.\,\ref{fig:brem only plots}(b) shows the energy distribution of the incident electrons, which had energy up to $\sim 500$ MeV, but large energy spread, as a result of emerging from an elongated 5 mm LWFA gas jet.  Since the 800 $\mu$m tantalum target significantly disrupted the electrons, preventing accurate on-shot measurement of their energy distribution, the black curve in this inset was obtained by averaging electron spectra of the 5 preceding shots \emph{without} the tantalum in place, while the grey shaded region represents their standard deviation. The average spectrum corresponds to a total of $3.4 \pm 1.1 \times 10^9$ electrons and average bunch energy 160 MeV, and was used for data analysis and Geant4 simulations.  

Blue dashed and solid black curves in Fig.\,\ref{fig:brem only plots}(a) show unfolded energy deposition profiles $D_i^{(calc)}$ for X-ray spectra 
based on the Born approximation (Eq.\,\ref{born nphot}a-b) and Kramers' law (Eq.\,\ref{brems model}), respectively. Fig.\,\ref{fig:brem only plots}(b) presents the corresponding best fit X-ray spectra. Red dotted curves in Fig.\,\ref{fig:brem only plots}(a) and (b) show the simulated energy deposition profile and photon spectrum, respectively. Unfolded and simulated energy deposition profiles are nearly indistinguishable from one another and fall within the 10\% relative uncertainty of the unfolding over the full range of the stack. Unfolded and simulated spectra similarly agree, with only small differences at the high energy limit (<20\%) between the two models. Table \ref{brem table} compares the bremsstrahlung beam parameters unfolded from the two models and obtained from the simulated spectrum.  The uncertainty in the number of photons in the simulated beam reflects the uncertainty in the number of electrons incident on the Ta target.

\begin{table}[htb!]
    \centering
        \begin{tabular}{ |c||c|c|c|c|c| }
        \hline
        & $E_{avg}$ (MeV) & $E_{cutoff}$ (MeV) & $N_{phot}$ & $N_{ph}/N_e$ & $E_{rad}/E_{bunch}$\\ 
        \hline
        Simulated  & 36 & $\sim500$  & $3.6 \pm 1.4 \times 10^{8}$ & $0.11 \pm .03$ & $0.024 \pm 0.008$  \\
         \hline
        Unf: Kramers  &  $35 \pm 4$ & $490 \pm 80$ & $4.2 \pm 0.8 \times 10^{8}$ & $0.12 \pm 0.04$ & $0.027 \pm 0.009$ \\
        \hline
        Unf: Born $N_{phot}$  & $36 \pm 5$ & $370 \pm 60$ & $4.1 \pm 0.8 \times 10^{8}$  & $0.12 \pm 0.04$ & $0.027 \pm 0.008$ \\
         \hline
        \end{tabular}
    \caption{Unfolded parameters for the two bremsstrahlung models and the simulated case including the average energy, cutoff energy, photon number, photon conversion efficiency $N_{ph}/N_e$ and energy conversion efficiency $E_{rad}/E_{bunch}$ over $100$ keV within the FWHM of the bremsstrahlung transverse energy profile for comparison.}
    \label{brem table}
\end{table}

\begin{figure}[!ht]
\centering
\includegraphics[width=.9\textwidth]{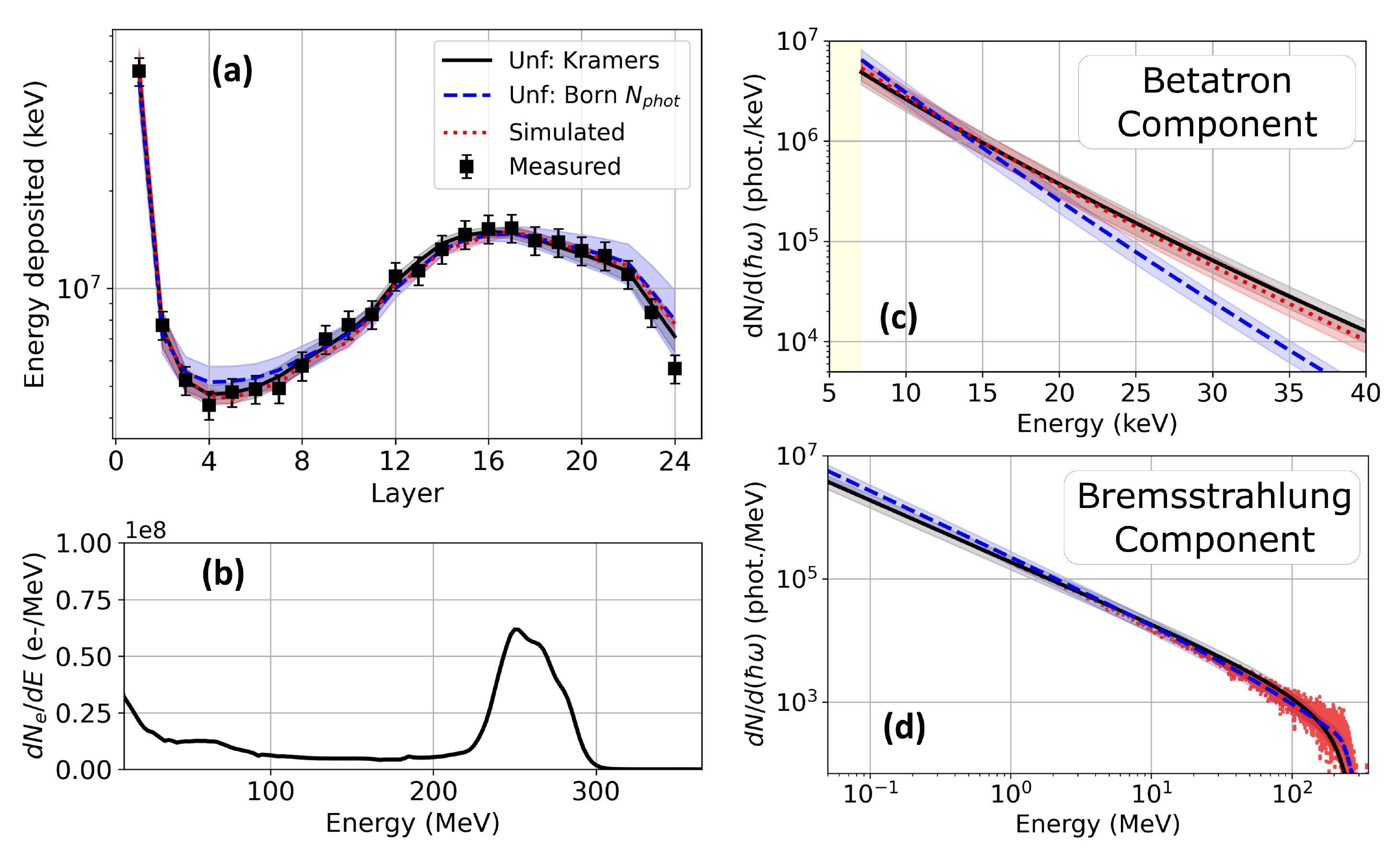}
\caption{\label{fig:brem beta plots} Combined betatron/bremsstrahlung X-ray results. (a) Measured energy deposition profile (black squares), unfolded profiles based on betatron radiation (Eq. 1) plus Kramers' law (black solid) and Born (blue dashed) bremsstrahlung models, and Geant4-simulated profile (red dashed) based on (b) the measured electron spectrum; (c) betatron and (d) bremsstrahlung spectra extracted from unfolded and simulated profiles in (a). Shaded regions denote unfolding error in (a), and uncertainty in absolute photon number in (c) and (d), as in Fig.\,4.}
\end{figure}

\vspace{0.2cm}
\noindent
\textbf{Betatron + bremsstrahlung X-rays}.
The 25 $\mu$m Kapton target was thick enough to generate detectable bremsstrahlung, yet thin enough to transmit most betatron radiation while negligibly perturbing the transverse spatial profile and energy spectrum of incident electrons. Data points (black squares) in Fig.\,\ref{fig:brem beta plots}(a) show a typical measured energy deposition profile $D_i^{(meas)}$ ($1 \leq i \leq 24$) using this target.  

%
%

Betatron radiation deposited most of the energy in layers 1-2, bremsstrahlung most of the energy in layers 5-24, while the two sources deposited comparable energy in intermediate layers 3-4. Because betatron and bremsstrahlung energy deposition profiles overlapped minimally, we analyzed and simulated each separately using models described in the previous two sections. We then unfolded the complete profile in one shot with the help of a single additional parameter describing their overall relative amplitude.  For data in Fig.\,\ref{fig:brem beta plots}(a), electrons originated from a 3-mm-long LWFA gas jet, yielding the energy spectrum with quasi-monoenergetic peak at $\sim$260 MeV shown in Fig. \ref{fig:brem beta plots}(b), which we measured on the same shot as the X-ray energy deposition profile. 

Black solid (blue dashed) curves in Fig.\,\ref{fig:brem beta plots}(a) represent unfolded deposited energy profiles based on Eq.\,1 for betatron radiation, on Kramers' Law (Born cross-section) for bremsstrahlung, and on an overall betatron/bremsstrahlung amplitude ratio parameter. We gave the Born cross-section model the form of Eq.\,\ref{born xsec main} (rather than 3), since electrons lose negligible energy in the thin target. Both fitted curves fall within experimental error bars throughout the detector stack, and differ noticeably from each other only in layers $4 < i < 7$.  Fig.\,\ref{fig:brem beta plots}(c) and (d) show corresponding betatron and bremsstrahlung spectra, respectively, while the last two rows of Table \ref{brem beta table} list unfolded model parameters for betatron radiation and bremsstrahlung. The $30\%$ difference in betatron parameters $E_c$ and $N_{phot}$ result from compensating for the difference between the two bremsstrahlung models in intermediate layers $4 < i < 7$.   

\begin{table}[ht!]
    \centering
        \begin{tabular}{ |c||c|c|c|c| }
        \hline
        & \multicolumn{2}{c|}{Betatron parameters} & \multicolumn{2}{c|}{Bremsstrahlung parameters}\\
        \hline
                        & $E_c$ (keV)   & $N_{phot}$ & $E_{avg}$ (MeV) & $N_{phot}$\\ 
        \hline
        Simulated       & $11 \pm 2$    &  $4.2 \pm 0.8 \times 10^{7}$ & 16.5 & $1.6 \times 10^{6}$\\
        \hline
        Unf: Kramers    & $12 \pm 2$    &  $3.8 \pm 0.7 \times 10^{7}$ & $15 \pm 2$ & $1.7 \pm 0.3 \times 10^{6}$ \\
         \hline
        Unf: Born x-sec & $9 \pm 2$     &  $4.8 \pm 1.0 \times 10^{7}$ & $12 \pm 3$ & $2.0 \pm 0.4 \times 10^{6}$ \\
         \hline
        \end{tabular}
    \caption{Betatron and bremsstrahlung X-ray parameters resulting from two model-based reconstructions of the combined energy deposition profile in the calorimeter, and from Geant4 simulation of the bremsstrahlung component. Betatron parameters include critical energy $E_c$, number of photons $N_{phot}$ with energy $> 7$ keV.  Bremsstrahlung parameters include average energy $E_{avg}$ and number of photons within the FWHM of the recorded calorimeter signal.}
    \label{brem beta table}
\end{table}

The red dotted line in Fig.\,\ref{fig:brem beta plots}(a) represents the ``simulated'' energy deposition profile. To obtain this curve, we first generated the bremsstrahlung part of the energy deposition profile in Geant4 using the measured electron spectrum [Fig.\,\ref{fig:brem beta plots}(b)], and scaled it vertically to match the measured energy deposition $D_i^{(meas)}$ in layers 8 through 24. We then used the remaining energy in the stack to unfold the betatron contribution based on Eq.\,1. The simulated profile also falls within experimental error bars throughout the stack, and nearly overlaps the unfolded ``Kramers'' profile. Likewise, the corresponding simulated spectra [red dotted curves in Fig.\,\ref{fig:brem beta plots}(c) and (d)] and model parameters (Table \ref{brem beta table}, third row from bottom) closely match their unfolded ``Kramers'' model counterparts.  Within uncertainty, we observed the same number $N_{phot}$ of betatron photons as from the pure betatron source. On the other hand, we observe 300 times fewer bremsstrahlung photons per electron from the thin Kapton target (Table \ref{brem beta table}) than from the thick tantalum target (Table \ref{brem table}).
 



\textbf{ICS X-rays.}  The ICS radiation spectrum is derived \cite{Esarey1993NonlinearPlasmas} from Li\'{e}nard-Wiechert potentials of accelerating electrons undergoing sinusoidal undulations in the electric field of a counter-propagating laser pulse. When the laser strength parameter $a_0 = 0.85 \lambda_0 (\mu \text{m}) \sqrt{I_0 (10^{18}~ \text{W/cm}^2)}$, analogous to a wiggler parameter, is much less than unity, radiation is generated in a forward directed cone at the Doppler-upshifted fundamental frequency\cite{Esarey1993NonlinearPlasmas} $4\gamma_e^2 \omega_0 /(1+a_0^2/2+\theta^2\gamma_e^2)$. Here, $\omega_0$ is the central laser frequency (and $\hbar \omega_0$ = 1.55 eV the central photon energy) and $\theta$ the angle of observation relative to the electron propagation direction. Assuming $a_0 \ll 1$ and given an electron spectrum $N_e f(\gamma)$, the energy radiated per unit $\hbar\omega$ can be calculated:

\begin{equation}\label{dIdhw}
    \begin{split}
       \frac{d\mathcal{E}_x}{d(\hbar \omega)} & = 2 \pi \alpha N_e N_0^2 a_0^2\int_0^{\theta_{max}} \sin\theta d\theta \int d\gamma f(\gamma) \gamma^2 \left( \frac{1 + \gamma^4 \theta^4 }{\left( 1 + \gamma^2 \theta^2 \right)^4}\right) Res(k,k_0)
    \end{split}
\end{equation}

\noindent
Here, $Res(k,k_0)$ is sharply peaked at the resonant frequency. This integration can take additional time and requires knowledge of the electron spectrum $N_e f(\gamma)$. For a peaked electron spectrum with relative energy spread $\sigma_\gamma/\gamma_e \approx 0.1$, the angle-integrated ICS spectrum can be approximated by a Gaussian function with mean photon energy $E_x$ and variance $\sigma_{E_x}$ (see Supplementary Material): 

\begin{equation}\label{ICS model}
\frac{d\mathcal{E}_x}{d(\hbar \omega)} \propto \hbar\omega \frac{dN}{d(\hbar\omega)} \propto C\exp\left(-\frac{(\hbar \omega-E_x)^2}{2 \sigma_{E_x}^2}\right).
\end{equation}

\vspace{0.2cm}
\noindent
Here, we express the spectral amplitude in terms of integrated X-ray pulse energy $\mathcal{E}_x = N\hbar\omega$ in order to retain a pure Gaussian function on the right-hand side.  The parameters $E_x$ and $\sigma_{E_x}$ must satisfy two physical constraints: (\textit{i}) $E_x$ cannot exceed $4 \gamma_{e}^2 \hbar \omega_0$; 
(\textit{ii}) $\sigma_{E_x}/E_x$ must exceed the relative energy spread of the electron bunch, i.e. $\sigma_{E_x}/E_x > \sigma_\gamma/\gamma$. The values of $E_x$ and $\sigma_{E_x}$ extracted from data analysis can then help to diagnose a variety of physical effects involved in ICS with a plasma mirror, e.g. redshift of laser photon frequency $\omega_0$ during LWFA, which decreases $E_x$; laser frequency broadening (here, $\sigma_{E_L}/E_L \approx 0.1$ or larger), electron energy spread (here, $\sigma_\gamma/\gamma \approx .065$) and angular divergence (here, $\sigma_\theta \approx 1/\gamma_e$), and non-linear scattering (generation of harmonics) \cite{Jochmann2013HighBackscattering,Kramer2018MakingApplications}, all of which contribute in quadrature to $\sigma_{E_x}$. 
Given the values of $\sigma_{E_L}/E_L$, $\sigma_\gamma/\gamma$ and $\sigma_\theta$ cited above, we constrain $\sigma_{E_x}/E_x$ to a practical lower bound of $0.35$ during unfolding.  

\normalsize
To generate ICS radiation, we used the thin Kapton foil to minimize bremsstrahlung, and placed it only $0 < z < 0.5$ cm from the LWFA exit to ensure strong retro-reflection of the spent LWFA drive pulse via plasma mirroring, thereby maximizing ICS.  Nevertheless betatron radiation from the LWFA leaked through the foil, and electrons from the LWFA generated some bremsstrahlung on passing through it. 
To remove the bremsstrahlung and betatron component, we took advantage of our ability, demonstrated in the preceding sections, to simulate the bremsstrahlung and betatron energy deposition profiles quantitatively and scale it to each shot based on the electron charge and average energy. 
We then subtracted this from the measured profile, leaving us with a pure ICS profile 
\textit{only}. The ratio of energy in the scaled bremsstrahlung/betatron profile to the total measured energy is $\sim11\%$ for both shots and agrees with independent scintillator based measurements of the relative contributions \cite{Hannasch2021NonlinearMirror}. The uncertainty in the final background subtracted ICS energy deposition profile incorporates the combined uncertainty in the measured data (10\% relative uncertainty) and the scaled bremsstrahlung/betatron uncertainty which we estimate has an increased relative uncertainty of 15\%. Thus, the final relative uncertainty is not constant for all layers and is higher for layers most affected by the subtraction procedure (layers 1 and 10-20). To include this modified uncertainty, the least squares optimization includes the relative uncertainty as a weighting for the unfolding. We will hereafter refer to the scaled bremsstrahlung/betatron profile as the background and the final ICS energy deposition profile after the subtraction procedure as the background-subtracted ICS data.

%
%
\begin{figure}[!ht]
    \centering
    \includegraphics[width=.95\textwidth]{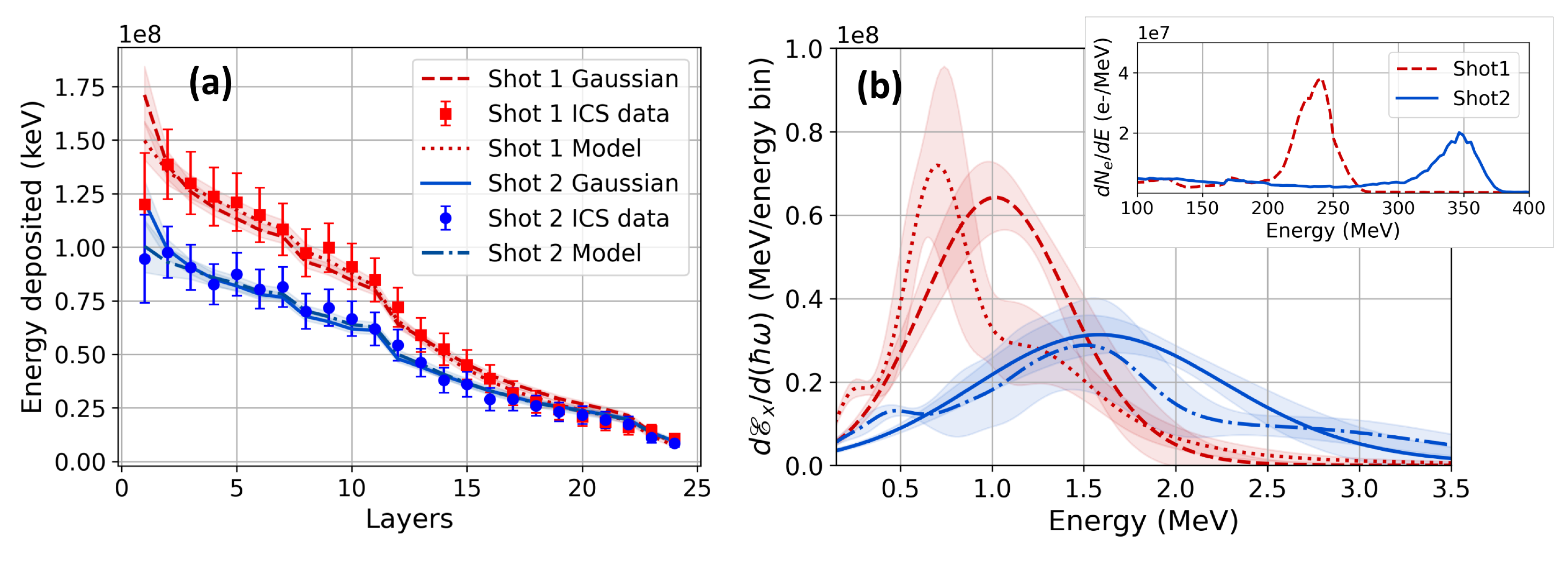}
    \caption{\label{fig:ics plots} ICS X-ray results. (a) Data points:  Background-subtracted ICS energy deposition profiles for two separate shots with peak electron energies $E_e = 236$ MeV (red squares) and 345 MeV (blue circles). Curves:  Best-fit reconstructed energy deposition profiles for ICS generated by 236 MeV electrons assuming a Gaussian spectrum (red dashed) and radiation model (red dotted) and 345 MeV electrons assuming a Gaussian spectrum (blue solid) and radiation model (blue dot-dashed). (b) Corresponding unfolded ICS spectra. Red and blue shading:  reconstruction uncertainty. Inset:  Electron spectra for the two shots.}
\end{figure}

Red and blue data points in Fig.\,\ref{fig:ics plots}(a) show background-subtracted energy deposition profiles of ICS generated on two separate shots, for which electron bunches had peak energy $236 \pm 14$ MeV ($\gamma_e = 462)$ and $345\pm13$ MeV ($\gamma_e = 675$), respectively [see red dashed and blue solid curves in the inset of Fig.\,\ref{fig:ics plots}(b)]. We achieved the lower and higher electron energies by tuning plasma density to $n_e = 4 \times 10^{18}$ $cm^{-3}$ and $6 \times 10^{18}$ $cm^{-3}$, respectively.  Red dashed and blue solid curves in Fig.\,\ref{fig:ics plots}(a) show best-fit energy deposition profiles from the unfolding process; corresponding curves in the main panel of Fig.\,\ref{fig:ics plots}(b) show unfolded ICS spectra of the form of Eq.\,\eqref{ICS model}. Red and blue shading around both pairs of curves represents unfolding uncertainty.  Table \ref{ics metrics table} lists ICS X-ray parameters $E_x$ and $\sigma_{E_x}$ of the unfolded spectra,
along with corresponding electron parameters for each shot. The unfolded $E_x$ values for the two shots stand in the ratio $E_x^{(236)}/E_x^{(345)} = 0.63 \pm 0.13$, whereas the expected $\gamma_e^2$ scaling of $E_x$ would yield a ratio $0.47 \pm 0.09$, assuming identical laser frequency $\omega_0$ on both shots.  While these ratios agree within the combined stated uncertainty, a possible reason for the discrepancy is that the laser pulse driving the denser plasma experienced a larger redshift, thus shifting the more energetic X-ray peak to lower energy.

\begin{table}[htb!]
    \centering
        \begin{tabular}{ |c||c|c|c|c|c| }
        \hline
        &\multicolumn{2}{c|}{Electron parameters} & \multicolumn{3}{c|}{Unfolded ICS parameters}\\
        \hline
        & $E_{pk}$ (MeV) & $N_{e}$ (>150 MeV) & $E_{x}$ (keV) & E spread ($\sigma_{E_x}$) & $N_{phot}$ (FWHM)\\ 
        \hline
        Shot 1  &  $236 \pm  14$ & $1.5 \times 10^{9}$ & $1040 \pm 90$ & $410 \pm 50$ & $8.2 \pm 2.0 \times 10^7$\\
        \hline
        Shot 2  & $345 \pm  14$ & $1.3 \times 10^{9}$ & $1640 \pm 190$ & $720 \pm 140$ & $4.9 \pm 1.0 \times 10^7$\\
        \hline
        \end{tabular}
    \caption{Electron parameters (left columns) and unfolded ICS X-ray parameters (right columns) based on Eq.\,\eqref{ICS model}, for two shots producing different peak energies $E_e$ and numbers $N_e$ of quasi-monoenergetic accelerated electrons.}
    \label{ics metrics table}
\end{table}

Table \ref{ics metrics table} presents the statistical average and standard deviation for $E_x$ and $\sigma_{E_x}$ for each shot. The two unfolded peaks are separated by more than their combined standard deviation and the unfolded value of $E_x$ for one peak falls outside of the FWHM of the second peak for 100\% of trials. We estimate a resolution of the unfolded $E_x$ of 100-200 keV for peak X-ray sources in the energy range of 500 keV to 2 MeV.

Simulations of the ICS spectrum require a good understanding of the 3-D laser intensity and the 6-D electron phase space to get accurate results of the farfield radiation spectrum \cite{Kramer2018MakingApplications}. However, the use of a plasma mirror makes it difficult to know the exact intensity and spectrum of the scattering laser pulse. Instead, a radiation model based on theory from Esarey \textit{et al.} (1993) \cite{Esarey1993NonlinearPlasmas} can be used to calculate the anticipated spectral shape generated by the measured electron spectrum scattering from a laser pulse of central frequency $\omega_0$ and laser strength parameter $a_0$ (see Supplementary Material). The calculation integrates over observation angles that would contribute to signal in the stack and assumes the central frequency of the scattering laser can be redshifted by a percent of the original e.g. $\omega_{scatter} = RS \omega_0$ where $RS \leq 1$. Calculations assuming several different values of $a_0$ in the range $0.1 \leq a_0 \leq 1.3$ were performed and the spectra resulting from other $a_0$ values within these bounds can be interpolated to provide a set of solutions to compare with the Gaussian model. 

An unfolding based on this radiation model finds $RS = 0.65 \pm 0.1~(0.6 \pm 0.1)$ and $a_0=0.55 \pm 0.2~(0.48 \pm 0.12)$ to be the values that best fit the measured energy profile $D_i^{(meas)}$ ($1\leq i \leq 24$) for shot 1 (shot 2). Fig.\,\ref{fig:ics plots}(a) shows the calculated energy deposition profiles $D_i^{(calc)}$ (dotted and dash-dotted curves) based on the best fit values of RS and $a_0$ for the radiation model. The corresponding spectra from the model are shown in Fig.\,\ref{fig:ics plots}(b) as dotted and dash-dotted curves with shading corresponding to the uncertainty of the unfolding. The goodness of fit defined by the fitness function $F(\bar{p})$ (see Methods) is $\sim 3 \times$ smaller for the radiation model that incorporates the electron spectrum compared with the Gaussian assumption. 
Moreover, the values of $a_0$ agree to within combined uncertainty with independent estimates of the laser intensity 1 mm after the exit of the accelerator \cite{Hannasch2021NonlinearMirror}.

\section*{Discussion}

The methods for unfolding the incident photon spectrum presented here are capable of capturing the characteristic radiation parameters for LWFA sources such as the betatron critical energy, bremsstrahlung average energy and ICS peak energy. The main limitation of the stack calorimeter detector and \emph{any} algorithm used to unfold the incident photon spectrum is that uniqueness of the solution cannot be guaranteed. Here we obtained a robust range of solutions by assuming physical models for the shape of the spectrum and deriving uncertainties from an ensemble of such solutions generated by sampling the measured energy distribution over 100 iterations. An example of this range is the unfolding of bremsstrahlung spectra using two different models that yielded indistinguishable energy deposition profiles with a thick, high-Z target. This multiplicity of solutions indicates that increasing the complexity of the radiation models to account for more features does not necessarily improve the certainty in the reconstructions. We have shown that applying simple assumptions based on the physics of each source can narrow down the range of solutions significantly. Moreover, the comparison for each case to models that incorporate the measured electrons such as the simulated betatron or simulated bremsstrahlung are in agreement within the unfolded uncertainty in all cases. The ICS source is an example where a model that incorporates the electron spectrum can provide a better fit to the observed energy profile, but both unfolded energy profiles fell within the uncertainty of the measured profiles for each shot. Nonlinear ICS in which $a_0$ approaches and exceeds 1 is just such a case where a more complex model may be necessary to unfold the harmonics of the fundamental, $4 \gamma_e^2 E_L$. Additionally, unfolding the spectra of X-rays radiated by electrons with multiply-peaked energy distributions will require models that incorporate such distributions explicitly.

Currently, using a least squares optimization algorithm, each single parameter case (betatron and bremsstrahlung dominated) converge to the solutions presented here in $\sim 1$ s. The case of multiple parameters (bremsstrahlung + betatron and ICS dominated) converges in $\sim 10$ s on a lab grade laptop. These algorithms can easily be transferred to manycore processors since each unfolding is performed 100 times and each run is independent. These computations can be parallelized to reduce the unfolding time by a factor of 100 or more to $\leq 100$ ms. To achieve data acquisition rates commensurate with such computational speed, image plates will need to be replaced with prompt scintillators compatible with $\sim$10 Hz LWFA repetition rates \cite{Rusby2018NovelExperiments}. In this geometry, plastic scintillators or scintillating fiber arrays alternate with absorbing material and the side of the stack is imaged with a camera or can be connected directly to photomultiplier tubes (PMTs)\cite{Wurden1995Scintillating-fiberOperation}. The analysis to generate the measured energy deposition profile $D^{(meas)}$ can also be parallelized since the operations on image data are independent. Moreover, the transmission speed of data along Gigabit-ethernet or USB 3.0 cables is $\sim 5-10$ Gbps and can transfer typical image sizes of 5 Mb in $<10$ ms. Cameras can already operate at the necessary 100 fps for this application. The limiting factor on the speed of unfolding is most likely in the conversion of data to a format for computation on a manycore processor. In total, current technology would allow a prompt scintillator based stack to operate at a minimum of 0.1 to 2 Hz, providing a method for actively unfolding spectra during LWFA experiments where emitted radiation provides a metric for the acceleration process e.g., enhanced betatron radiation from direct laser acceleration (DLA) or higher order harmonics in non-linear ICS.  

We have presented a set of unfolded secondary X-ray spectra spanning over 4 orders of magnitude in energy from LWFA accelerated electrons with energies between 230 MeV and 550 MeV. The LWFA and target geometry can be tuned to generate betatron, bremsstrahlung or ICS dominated sources as well as a regime in which both betatron and bremsstrahlung contribute to the stack. We present unfolding of betatron radiation with a critical energy of $14\pm1.5$ keV and betatron radius of $1.0\pm0.1~\mu$m which are compared with independent measurements using a X-ray sensitive CCD and simulations from CLARA2. Bremsstrahlung from an 800 $\mu$m tantalum target is unfolded with an average energy of $35\pm 4$ MeV and $4.2\pm0.8\times10^8$ photons within the FWHM and is compared with Geant4 simulations. Thin-target bremsstrahlung from $25~\mu$m of Kapton includes contribution from both betatron and bremsstrahlung and the unfolded critical energy of the betatron source is $12\pm3$ keV and the average bremsstrahlung energy is $14\pm 3$ MeV, spanning 3 orders of magnitude in a single shot. Finally, ICS dominated radiation from electron bunches with different peak energies was unfolded to observe a shift in peak ICS energy from $1060\pm90$ keV to $1.64\pm0.19$ MeV and a total of $8.2\pm0.2\times10^{7}$ and $4.9\pm1.0\times10^{7}$ photons in the FWHM, respectively. The ICS shots were compared with an electron dependent model that unfolded a value for $a_0$ of $0.55\pm0.2$ and $0.48\pm0.12$ and a relative redshift (RS = $\omega_L/\omega_0$) in the laser central frequency of $0.65\pm0.1$ and $0.6\pm0.1$. The stack calorimeter is less sensitive to background and has a higher signal to noise ratio for the energy ranges presented here than similar spectrometers that rely on a Compton converter \cite{Singh2018CompactExperiments,Tiwari2019GradientRange} or Ross filter pairs \cite{Ross1928AX-Radiations,Khutoretsky1995DesignKeV}. Furthermore, stack calorimeters are compact in size, making them ideal detectors for characterizing X-ray sources from a variety of laser systems.

\section*{Methods}


\textbf{Laser wakefield electron acceleration.}  Pulses of $30$\,fs duration, $800$\,nm center wavelength from the DRACO laser at Helmholtz-Zentrum Dresden-Rossendorf (HZDR) \cite{Couperus2017DemonstrationAccelerator,Schramm2017FirstDresden} were focused to spot size 20 $\mu$m (FWHM) with typical energy 2\,J onto the entrance plane of a 3-mm or 5-mm-long He gas jet doped with 1\% Nitrogen.  
The laser pulse fully ionized the helium, creating plasma of electron density in the range $4 < n_e < 6 \times 10^{18}$\,cm$^{-3}$, and drove a LWFA in the self-truncated ionization-injection regime \cite{Couperus2017DemonstrationAccelerator,Mirzaie2015DemonstrationBeams}.  A magnetic electron spectrometer \cite{Schramm2017FirstDresden,Kurz2018CalibrationDetermination} with its entrance plane at $z = 30$\,cm downstream of the gas jet exit determined the electron energy distribution for each shot.  Fig.\,\ref{fig:experimental schematic}(b) shows an example for the 3-mm jet.  The spectrum consists of a quasi-monoenergetic peak with central energy in the range $200 < E_e < 350$\,MeV (Lorentz factor $400 < \gamma_e < 700$), energy spread $\sim$20\,MeV (FWHM), rms divergence 2 mrad and charge in the range $200 < Q < 300$\,pC, which is responsible for most X-ray production, and a weak poly-energetic, low-energy background.  The 5-mm jet yielded electrons with energy up to 550\,MeV, with a stronger poly-energetic background.  

\vspace{0.2cm}
\noindent
\textbf{X-ray spectral reconstruction.}  We write the integrated energy deposited in layer $i$ of the calorimeter as a vector with components $D_i (i=1,2,...,24)$.  We wish to reconstruct from this the spectrum $dN/d(\hbar\omega)$ of incident X-rays, which we discretize as a vector $dN_j/d(\hbar\omega)$ describing the number of photons in bin \textit{j} of energy $\hbar \omega_j$ and width $d(\hbar\omega_j)$.  A stack response matrix $R_{ij}$ describes the energy deposited in layer \textit{i} by photon of energy $\hbar \omega_j$ and relates $D_i$ to $dN/d(\hbar\omega_j)$ via \cite{Horst2015AInteraction}:


\begin{equation}\label{dose calculation}
    D_i \approx \sum_{j=1}^{N} \frac{dN_j}{d(\hbar \omega)} R_{ij}  ~d(\hbar \omega_j),
\end{equation}

\noindent
where the sum is over the number of energy bins, $N$. Here, $N \approx 1600$, with $d(\hbar\omega_j) = 1$\,keV for $5$\,keV $< \hbar\omega_j < 200$\,keV, $d(\hbar\omega_j) = 20$\,keV for $200$\,keV $< \hbar\omega_j < 10$\,MeV, $d(\hbar\omega_j) = 250$\,keV for $10$\,MeV $<\hbar\omega_j < 200$\,MeV, $d(\hbar\omega_j) = 1$\,MeV for $200$\,MeV $< \hbar\omega_j < 400$\,MeV and $d(\hbar\omega_j) = 5$\,MeV for $400$\,MeV $< \hbar\omega_j < 600$\,MeV.  We generate $R_{ij}$ by simulating energy deposition in the stack's absorbers and IPs by mono-energetic photon beams of different $\hbar\omega_j$ using Geant4 \cite{Agostinelli2003GEANT4Toolkit}. A reconstruction begins with an initial guess of $\frac{dN_j}{d(\hbar\omega)}(\hbar\omega_j,\bar{p})$, which here is constrained to take the form of a physics-based analytic function of $\hbar \omega_j$, typically including a small set $\bar{p}$ of fit parameters, describing betatron, ICS or bremsstrahlung radiation, or a combination of them.  Specific functions used for each type of X-ray source are presented in the Results. Knowledge of the presence and location of PMs and converters, and other experimental parameters, is critical in choosing appropriate functions. The most accurate models take the measured electron spectrum (Fig.\,\ref{fig:experimental schematic}(b)) specifically into account.  However, models that do not depend explicitly on the electron spectrum are also useful for rapid, albeit approximate, results.  In either case, a forward calculation using equation\,\ref{dose calculation} generates a first-generation $D^{(calc)}_i$, which is compared to the measured energy distribution $D^{(meas)}_i$. A fitness function

\begin{equation}\label{fitness func}
    F(\bar{p}) = \sum_{i=1}^{n} (D^{(meas)}_i - D^{(calc)}_i)^2 = \sum_{i=1}^{n} \left(D^{(meas)}_i - \left[\sum_{j=1}^{N} \frac{dN_j}{d(\hbar \omega)} R_{ij}  ~d(\hbar \omega_j) \right] \right)^2,
\end{equation}

\noindent
i.e. the sum of squared residuals between the calculated and measured energy, then evaluates the goodness of fit where, $n$ denotes the number of layers.  In subsequent iterations, $\frac{dN_j}{d(\hbar\omega)}(\hbar\omega_j,\bar{p})$ is varied in an effort to minimize $F(\bar{p})$. Here, we unfold the spectral shape, not the absolute value, of $\frac{dN_j}{d(\hbar\omega)}(\hbar\omega_j,\bar{p})$, by fitting to the energy distribution $D_i$ normalized to total deposited energy $\sum_{i=1}^n D_i$. 
The overall scaling is reintroduced after the completed unfolding to account for the total energy in the beam (see Supplementary Material for stack calibration). As in solving any complex inverse problem with incomplete information, convergence of the iterative procedure and uniqueness of any best fit solution cannot be guaranteed.  Thus thorough tests of the sensitivity of results to initial guesses, awareness of experimental conditions, liberal use of physical constraints on the form of solutions and accurate evaluation of error are essential to achieving reliable results.

\vspace{0.2cm}
\noindent
\textbf{Analyzing stack data.}  For each IP layer in the stack the deposited energy is integrated within the FWHM of the incident beam to determine the measured energy distribution in the stack, $D_i^{(meas)}$ (plotted in Fig.\,\ref{fig:Dose comparison}). The divergence of the incident photon beams is found by averaging the divergence in each layer over the relevant layers for each X-ray source. The betatron divergence is found using only layer 1, while the divergence of the bremsstrahlung and ICS sources is averaged over layers 5-18 to avoid an overestimation caused by betatron contributions or scattering in the high Z layers. Table \ref{dose analysis} compiles the measured beam divergence for each presented case, the radius of integration for $D_i^{(meas)}$ an the resulting energy deposited. In the case of the betatron + bremsstrahlung X-rays from a 25 $\mu$m-thick Kapton target, the energy deposition profile $D_i^{(meas)}$ is integrated over a radius corresponding to the \textit{bremsstrahlung} HWHM of $3.4 \pm 0.1$ for unfolding both sources. The unfolded betatron spectrum is then scaled to the energy integrated within a radius of 7 mrad corresponding to the betatron HWHM for direct comparison with the betatron dominated case. 

\begin{table}[htb!]
    \centering
        \begin{tabular}{ |c||c|c|c| }
        \hline
        & Beam HWHM (mrad) & Integration radius (mrad) & Total energy deposited (keV) \\ 
        \hline
        Betatron  & $7.7 \pm 0.5$ & 7.7 &  $2.8 \pm 0.6 \times 10^8$ \\
         \hline
        800 $\mu$m Ta bremsstrahlung  & $5.7 \pm 0.2$ & 5.7 & $1.0 \pm 0.2 \times 10^{11}$  \\
        \hline
        \makecell{25 $\mu$m Kapton \\ bremsstrahlung (betatron)} & $3.4 \pm 0.1$ ($7\pm2$)& 3.4 (7) & $2.7 \pm 0.5 \times 10^8$   \\
        \hline
        ICS shot 1  & $4.5 \pm 0.2$ & 4.5 &  $1.8 \pm 0.3 \times 10^9$ \\
        \hline
        ICS shot 2  & $3.5 \pm 0.3$ & 3.5 &  $1.4 \pm 0.3 \times 10^9$ \\
        \hline
        \end{tabular}
    \caption{Compiled divergence, integration radius and integrated energy for each X-ray source presented in the text. The energy deposited for the 25 $\mu$m Kapton bremsstrahlung case is integrated within the HWHM of the bremsstrahlung beam and then scaled to the betatron beam HWHM (shown in parentheses) after unfolding.}
    \label{dose analysis}
\end{table}

\vspace{0.2cm}
\noindent
\textbf{Error management.}  Uncertainty and error in measured energy deposition distribution $D^{(meas)}_i$ propagate into uncertainties and errors in recovered X-ray spectra $dN_j/d(\hbar\omega)$, and must therefore be carefully evaluated.   Calibration of IP sensitivity and scanner introduce uncertainty of order $\pm20\%$ into the absolute value of measured energy (see Supplementary Material).  Variability of the fading rate of IP luminescence (typically $0.78\pm0.03$ when scanned 10-15 minutes after exposure) \cite{Tanaka2005CalibrationSpectrometer} introduces additional uncertainty.  Fortunately, most of this uncertainty affects only overall energy deposited and absolute energy of the beam, not the \emph{shape} of the energy deposition from which $dN_j/d(\hbar\omega)$ is unfolded. Nevertheless, layer-dependent errors arise when IPs with different ages, manufacturing and usage histories, and distributions of defects are mixed together in a stack.  Repeated exposures of the same IP yield up to $\sim5\%$ rms variation in recorded PSL \cite{Rosenberg2019Image-plateKeV}.  Based on this measurement, we estimated $\sim10\%$ rms variations within a stack, to take into account age and sensitivity difference among different IPs.  Such variations introduce uncertainty into the normalized shape of the energy distribution, and hence into parameters of the unfolded spectrum. To take this into account, we randomly generate a normal distribution of synthetic energy profiles $D_i^{(syn)}$ with standard deviation of $10\%$ around the measured profile $D_i^{(meas)}$. This ensemble of synthetic energy profiles then becomes the target for unfolding.  Each iteration uses one distribution from the ensemble as a target; the procedure is repeated $\sim~$100 times using different $D_i^{(syn)}$ to obtain an equivalent ensemble average and standard deviation for the spectrum $dN_j/d(\hbar\omega)$, and for a given model's parameter set $\bar{p}$.

\vspace{0.4cm}
\noindent
\textbf{\large{Data Availability.}} The data that support the plots within this article and other findings of this study are available from the corresponding author upon reasonable request.

\bibliography{references_static}

\section*{Acknowledgements}
U.\,Texas authors acknowledge support from U. S. Department of Energy grant DE-SC0011617, A.H. from the National Science Foundation Graduate Research Fellowship grant No. DGE-1610403 and M. C. D. from the Alexander von Humboldt Foundation.  HZDR authors acknowledge support from the Helmholtz Association under program Matter and Technology, topic Accelerator R \& D. 

\section*{Author contributions statement}

A.H., A.L.G., M.L., R.Z., J.C.C., O.Z., T.K., A.F., M.M. and A.I. conducted the experiments. A.H. and A.L.G analyzed the results. A.H., A.L.G. and A.K. performed the simulations and A.K. conducted the experiment that contributed to data in Fig.\ref{fig:betatron only plots}(b). L.N. helped with radioactive sources. T.E.C., U.S., A.I and M.C.D. provided overall supervision of the project.  All authors reviewed the manuscript.

\section*{Additional information}
 
\noindent
\textbf{Competing interests:} The authors declare no competing interests.

\end{document}


\section*{Supplementary material}

\subsection*{Stack composition, response and calibration}
\begin{figure}[!ht]
    \centering
    \includegraphics[width=.75\textwidth]{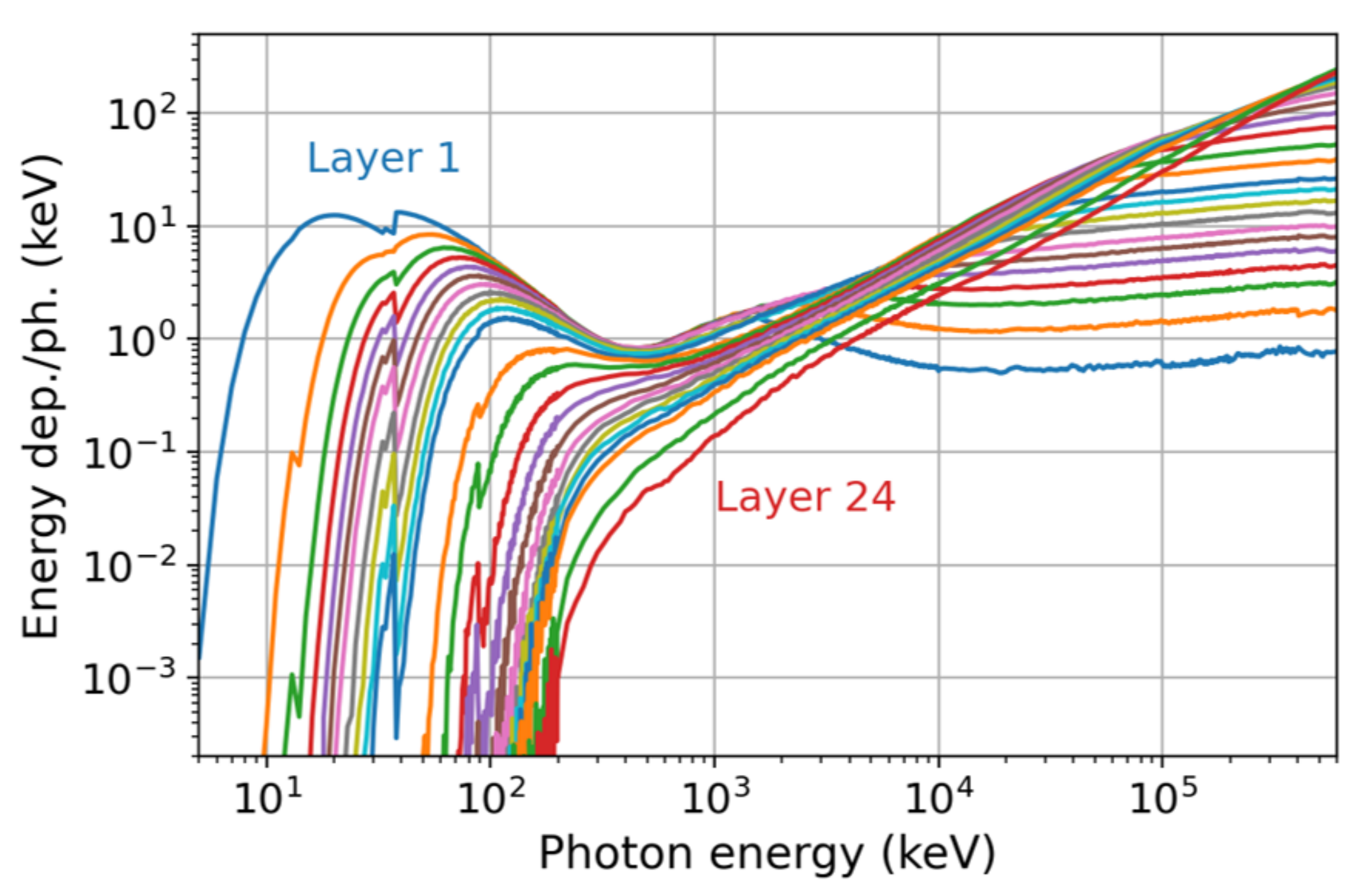}
    \caption{\label{fig:response curves}Response curves in energy deposited per photon for layers 1 through 24 and from 5 keV to 600 MeV in varying step sizes.}
\end{figure}

The stack design is composed of alternating absorbing materials of varying Z and thickness including PMMA, Aluminum, Brass and Steel as outlined in table\,\ref{stack materials}. Fuji BAS-MS image plates (IP) are placed behind each absorbing layer to record the energy deposited by ionizing radiation. IPs have a large dynamic range and are sensitive to ionizing particles, recording the two-dimensional energy deposition of a radiation source either directly from the X-rays or from secondary particles such as electrons or positrons \cite{Bonnet2013ResponseParticles}. The response curves in Fig.\,\ref{fig:response curves} are generated by simulating mono-energetic, non-divergent photon beams interacting with the stack in Geant4 and recording the energy deposited per photon in each IP. The simulated IPs have a composition based on Rabhi \textit{et al.} (2016) \cite{Rabhi2016CalibrationMeV}.  

\begin{table}[htb!]
    \centering
        \begin{tabular}{ |c|c|c||c|c|c| }
        \hline
        Layer & Thickness & Material & Layer & Thickness & Material \\
        \hline
        1 & 2 mm & PMMA & 13 & 2 mm & Brass \\ 
        \hline
        2 & 2 mm & PMMA & 14 & 3 mm & Brass \\
        \hline
        3 & 3 mm & PMMA & 15 & 3 mm & Brass \\
        \hline
        4 & 3 mm & PMMA & 16 & 3 mm & Brass \\ 
        \hline
        5 & 5 mm & PMMA & 17 & 3 mm & Brass \\
        \hline 
        6 & 5 mm & PMMA & 18 & 3 mm & Brass \\
        \hline 
        7 & 5 mm & PMMA & 19 & 3 mm & Steel \\
        \hline
        8 & 3 mm & Aluminum & 20 & 3 mm & Steel \\
        \hline
        9 & 3 mm & Aluminum & 21 & 3 mm & Steel \\
        \hline
        10 & 4 mm & Aluminum & 22 & 4 mm & Steel \\ 
        \hline
        11 & 4 mm & Aluminum & 23 & 13 mm & Steel \\
        \hline
        12 & 2 mm & Brass & 24 & 10 mm & Steel \\
        \hline
        \end{tabular}
    \caption{List of stack absorbing materials and thicknesses. Alternating BAS-MS image plates are placed in plastic packets after each absorbing layer to record the energy deposition by the incoming radiation. The design remained consistent for all radiation sources presented here.}
    \label{stack materials}
\end{table}

\begin{figure}[!ht]
    \centering
    \includegraphics[width=.75\textwidth]{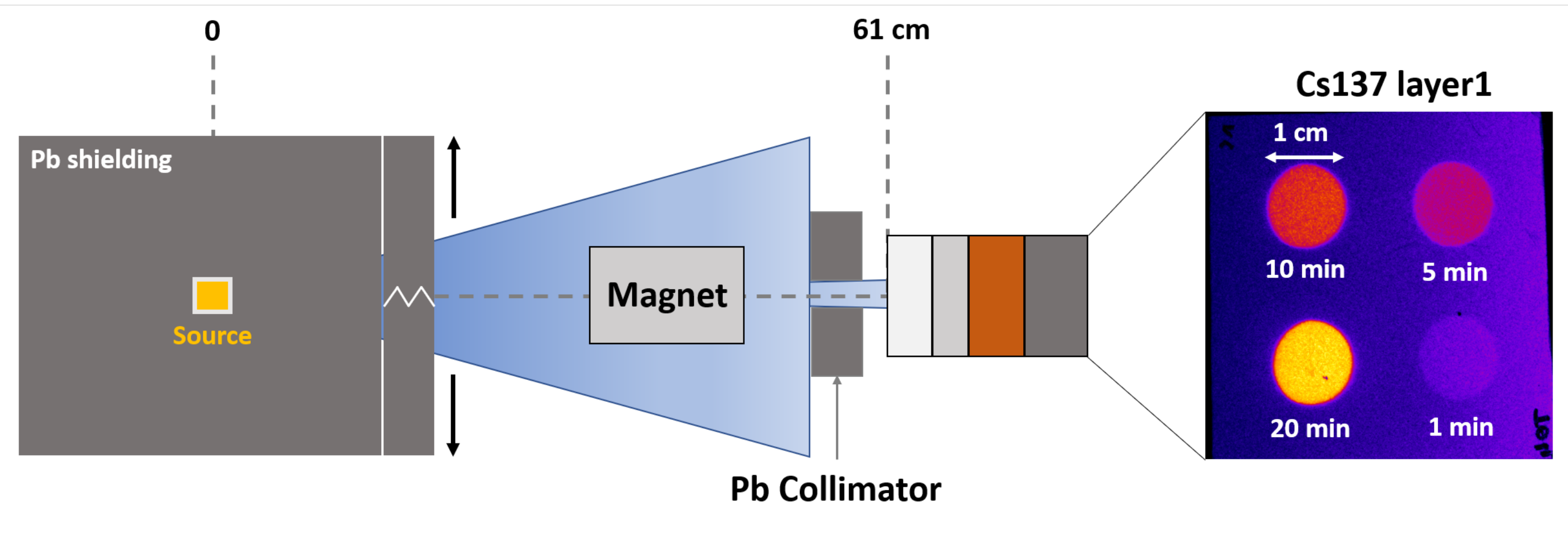}
    \caption{\label{fig:source setup}Calibrated radioactive sources are housed in a Pb shielded box with a 30$^\circ$ aperture. Exposure times of 1, 5, 10 and 20 minutes are performed with the stack placed 61 cm away from the source capsule. A magnet and 1.2 cm Pb collimator are placed prior to the stack to remove secondary electrons and make multiple exposures possible.}
\end{figure}

Calibration of the stack was performed using Cs137 and Co60 sources with activities of 9.25 GBq and 10 GBq with 10\% tolerance, respectively. The sources are 4 mm x 4 mm cylindrical capsules encased in $\sim 1.2$ mm of stainless steel and are housed in a lead shielded box with mechanical lead doors and a $30^\circ$ aperture. Fig.\,\ref{fig:source setup} shows the set up used for these measurements and includes a magnet to disperse low energy electrons and a Pb collimator with 1.2 cm diameter to isolate the characteristic X-rays. The stack calorimeter was placed after the Pb collimator for 1, 5, 10 and 20 minute exposures and then scanned $\sim$ 30 minutes after the 20 minute exposure. The energy deposition profile is found by integrating the signal within a circle of 1 cm diameter for each exposure and subtracting the BG from a region between each exposure (see Fig.\,\ref{fig:source setup}). A SpectroTRACER scintillator based spectrometer was also exposed to each source to provide a calibrated measurement of the spectrum and a BeO ceramic dosimeter measured the calibrated energy used to calculate the total flux of each source. 

\subsubsection*{Response validation}

\begin{figure}[!ht]
    \centering
    \includegraphics[width=.9\textwidth]{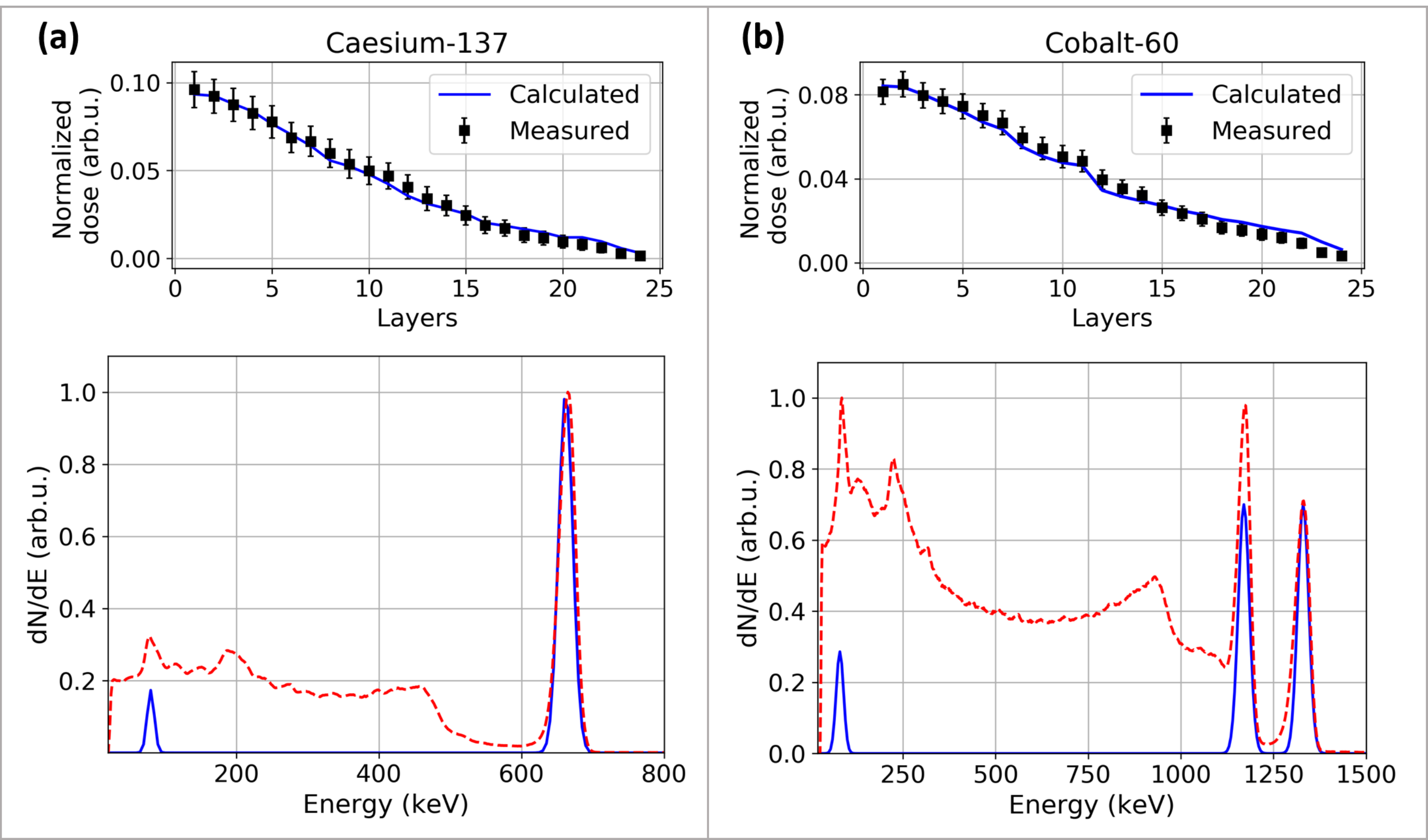}
    \caption{\label{fig:source plots}Top panels compare the 20 minute exposure after background subtraction (data points) and calculated energy profiles (blue) for (a) Cs137 and (b) Co60. The corresponding incident photon spectrum without the Compton continuum (blue solid) and with the Compton continuum from the SpectroTRACER spectrometer (red dashed) are compared in the bottom panels.}
\end{figure}

A general unfolding of the Cs137 and Co60 source spectra is not the goal for this manuscript. We aim only to provide validation that the response matrix is accurate given the incident photon spectrum is known or well-understood. The bottom panels in Fig.\,\ref{fig:source plots}(a) and (b) show the calibrated SpectroTRACER spectrum (red dashed curve) for Cs137 and Co60, respectively. The characteristic energy is clearly visible in each case at 662 keV for Cs137 and at 1.17 and 1.33 MeV for Co60. Other visible features include a fluorescence peak at $\sim 88$ keV from the Pb collimator as well as the Compton continuum resulting from characteristic X-rays scattering within the scintillator and then escaping. The scattered electrons in these events deposit the transferred energy in the scintillator which reads as a measured event in the SpectroTRACER output. These Compton scattering events similarly occur within the absorbing materials in the stack and deposit energy in the image plates. However, energy deposition by Compton electrons or $e^+ e^-$ pairs produced in the interaction of a $\gamma$-ray with the material are calculated in the Geant4 Electromagnetic process and model classes used for simulating the response matrix. Thus, the only source of energy deposition are by photons entering the stack from the characteristic X-rays for each source and the Pb fluorescence peak. This fluorescence peak can be estimated from the SpectroTRACER spectra and subsequently unfolded to optimize the fit [see bottom panels of Fig.\,\ref{fig:source plots}, blue solid curve].

Additional background from electron decay products generating bremsstrahlung can contribute to the stack profile as well, but simulations indicate that the amplitude of the background is lower by a factor $\sim2-5\times10^{-4}$ compared with the characteristic peaks due to re-absorption in the source and surrounding steel capsule. Fig.\,\ref{fig:source plots} (top panels) show that the agreement between the measured energy deposition profile (data points) and calculated energy deposition profile from the spectrum including only the characteristic peaks and Pb fluorescence peaks (blue solid curve) is within $\pm 10\%$. These examples indicate that the response of the stack to photons in the range of 0.1 - 1.3 MeV is accurate to within 10\%. 

\subsubsection*{Total flux}

A BeO detector \cite{Jahn2013TheApplications} was used to measure the calibrated dose for each source in combination with a magnet and Pb collimator. The flux was calculated assuming only photons at the characteristic energy for Cs137 and Co60 and is $1.6\times10^5$ and $3.7\times10^5 cm^{-2}s^{-1}$ respectively. The flux into the stack is also estimated by assuming only photons at the characteristic energies and allows us to calculate the scaling factor, $\alpha$, between PSL and energy for the image plates and scanner used in this experiment. This factor varies only mildly between the two sources, $\alpha_{Cs137} = 2.7\pm0.6\times10^{-4}$ PSL/keV and $\alpha_{Co60} = 3.0\pm0.5\times10^{-4}$ PSL/keV, and each are within error of the other. These values are smaller than similar calibrations performed in the field \cite{Bonnet2013ResponseParticles,Boutoux2015StudyElectrons} but this can be caused by a lower sensitivity of the scanner used for our experiments and a lack of recent re-calibration \cite{Zeil2010AbsoluteBunches}. For results presented here, we use $\alpha = 2.9\pm0.6\times10^{-4}$ PSL/keV to convert the PSL to energy deposition in the stack. 

\subsection*{CLARA2 simulations for betatron radiation}

CLARA is a \textbf{CLA}ssical \textbf{RA}diation code that calculates radiation spectra, $d^2I/d\omega d\Omega$, in the farfield from the Li\'{e}nard-Wiechert potentials \cite{Pausch2014HowFrameworks}. 

\begin{equation}\label{radiation equation}
\frac{d^2 I}{d\omega d\Omega} = \frac{q^2}{16 \pi \epsilon_0 c} \Big| \int_{-\infty}^{+\infty} \frac{\bar{n}\times\left[ \left( \bar{n} - \bar{\beta} \right) \times \dot{\bar{\beta}}\right]}{(1-\bar{\beta} \cdot \bar{n})} \cdot e^{i\omega(t-\bar{n}\cdot\bar{r}(t)/c)}\Big|^2
\end{equation}

The calculation takes the particle trajectories, $\bar{r}(t)$, and energy as inputs and is parallelized to run on large CPU clusters. To calculate the spectra from betatron emission the electron trajectories are found using a linear energy gain acceleration model  where the electrons experience sinusoidal oscillations with a time-dependent amplitude, $r_b(t)$, and frequency, $\omega_b(t)$.

\begin{equation}\label{e trajectory}
    x(t) = r_b(t)\cos(\phi_b(t))
\end{equation}

\noindent
Here, $x(t)$ is the 1-D particle path where 

\begin{equation}\label{initial cond}
    r_b(t) = r_{b0}\left(\frac{\gamma(t)}{\gamma_0}\right)^{-1/4} ~\text{and}~ \phi_b(t)=\int_0^t \omega_b(t) dt 
\end{equation}

\noindent
Both the oscillation amplitude and frequency depend on the time-dependent energy where $\omega_b(t) = \omega_p/\sqrt{2 \gamma(t)}$. To calculate the single electron trajectory, the energy gain is estimated to be linear \cite{Glinec2008DirectAccelerator}

\begin{equation}\label{energy gain}
    \gamma(t) = \gamma_0 + \frac{e E_z}{m_e c}t 
\end{equation}

\noindent
where $E_z$ is the accelerating field and $\gamma_0$ is related to the phase velocity of the bubble upon injection such that $\gamma_0 \approx \gamma_\phi = 1/\sqrt{1-(v_\phi/c)^2}\approx \omega_0/\sqrt{3}\omega_p$ \cite{Lu2007GeneratingRegime,Albert2013AngularAccelerator}. For the experimental shots presented here, $n_e = 5\times10^{18} ~cm^{-3}$ and $\gamma_0 \sim 10$. The final energy $\gamma_f$ is then extracted from the measured electron spectrum from the experiment.

To find the betatron radius for an electron bunch with an arbitrary distribution of $r_{b0}$ values, a set of electron trajectories is calculated using equations\,\eqref{e trajectory}-\eqref{energy gain} and input into CLARA2 to find the $r_{b0}$ dependent betatron spectra. The final betatron spectrum from the electron bunch can be found by weighting the set of CLARA2 spectra with the initial $r_{b0}$ distribution. 

\subsection*{Bremsstrahlung model approximations}

Calculations of the cross-section based on the Born approximation hold for cases when $Z/137 < v/c$ and applies to the scattering of relativistic electrons presented here \cite{Bethe1934OnElectrons}. The differential cross-section in photon energy $\hbar\omega$ (neglecting screening effects) from an electron of energy $E_0$ is

\begin{equation}\label{born xsec}
\left(\frac{d\sigma}{d(\hbar\omega)}\right)_{Born} = \frac{16}{3}  \frac{Z^2 r_e^2 \alpha}{\hbar\omega} \left( 1 - \frac{\hbar\omega}{E_0}+\frac{3\hbar^2\omega^2}{4 E_0^2}\right)\left[\ln\left(\frac{2E_0(E_0-\hbar\omega)}{m_e c^2\hbar\omega}\right)-\frac{1}{2}\right]
\end{equation}

\noindent
Here, Z is the charge of the scattering nucleus, $\alpha$ is the fine structure constant and $r_e$ is the classical electron radius. For a high Z target and relativistic electrons ($E_0>>137mc^2Z^{-1/3}$), screening of the nucleus by atomic electrons can have an energy dependent effect on the cross-section and a correction to equation\,\eqref{born xsec} is required. The cross-section when screening is considered can be written analytically as

\begin{equation}\label{screening xsec}
\left(\frac{d\sigma}{d(\hbar\omega)}\right)_{screening} = \frac{16}{3}  \frac{Z^2 r_e^2 \alpha}{\hbar\omega} \left[\left( 1 - \frac{\hbar\omega}{E_0}+\frac{3\hbar^2\omega^2}{4 E_0^2}\right)\ln\left(183  Z^{-1/3}\right)+ \frac{1}{9} \left(1-\frac{\hbar\omega}{E_0}\right)\right]
\end{equation}

For the screened case, the last term in the brackets contributes $\sim 3\%$ compared to the first term for tantalum ($Z=73$) and can be dropped for the approximation without affecting the relative shape of the spectrum. Then, the photons emitted per energy bin can be estimated by integrating the cross-section over the electron energy loss and by assuming that the electron loses energy to radiation at a rate proportional to its energy, $dE/dx = -E/L_0$ \cite{Bethe1934OnElectrons}. 

\[\frac{dN}{d(\hbar\omega)} = n N_e \int_{\hbar\omega}^{E_i} \frac{d\sigma}{d(\hbar\omega)} \frac{dE_0}{(-dE_0/dx)} = n N_e L_0 \int_{\hbar\omega}^{E_i} \frac{1}{E_0} \frac{d\sigma}{d(\hbar\omega)} dE_0 \]

\noindent
Here, $n$ is the target density, $L_0$ is the radiation length, $N_e$ the number of electrons and $E_i$ is the initial electron energy. Below a critical energy of $E_0 = 1600mc^2/Z$ the energy loss is dominated by collisions and the electron no longer contributes significantly to the photon spectrum. For tantalum, this energy is $\sim 11 MeV$ and would require that the highest energy electrons lose $\sim 98\%$ of their energy. This energy loss translates to a tantalum thickness of $4\times$ the radiation length, or 1.2 cm. We performed the bremsstrahlung experiments with 800 $\mu m$ tantalum target thickness and so we should not lose more than 20\% of the electron energy to radiation. This still requires the integration over the cross-section and the result is a piece-wise function to account for photon energies above and below the final electron energy. 

\begin{subequations}
\begin{align}
\frac{dN}{d(\hbar\omega)} = n N_e L_0 \int^{E_i}_{\hbar\omega} \frac{1}{E_0} \frac{d\sigma}{d(\hbar\omega)}dE_0 , \hspace{1.5cm} E_f\leq \hbar\omega\leq E_i
\\
\frac{dN}{d(\hbar\omega)} = n N_e L_0 \int^{E_i}_{E_f} \frac{1}{E_0} \frac{d\sigma}{d(\hbar\omega)}dE_0 , \hspace{1.5cm} \hbar\omega\ < E_f \leq  E_i 
\end{align}
\end{subequations}

\noindent
The final electron energy $E_f$ is calculated based on the target thickness $t$ and radiation length $L_0$ of the material as $E_f = E_0 \exp{(-t/L_0)}$, where $E_0$ is the initial electron energy. This integration can be done analytically if we assume the form of the cross-section for full screening and remains computationally efficient since the only input is the initial electron energy and target material information. The photon spectrum can be written analytically as  

\begin{subequations}
\begin{align}
\left(\frac{dN}{d(\hbar\omega)}\right)_{low} = \frac{C}{\hbar\omega} \left( \ln{\frac{E_0}{E_f}}+\hbar\omega \left(\frac{1}{E_0}-\frac{1}{E_f}\right) - \frac{3}{8} (\hbar\omega)^2 \left(\frac{1}{E_0^2}-\frac{1}{E_f^2}\right) \right), \hspace{1.5cm} \hbar\omega\ < E_f \leq  E_0  \\
\left(\frac{dN}{d(\hbar\omega)}\right)_{high} = \frac{C}{\hbar\omega} \left( \ln{\frac{E_0}{\hbar\omega}}+\hbar\omega \left(\frac{1}{E_0}-\frac{1}{\hbar\omega}\right) - \frac{3}{8} (\hbar\omega)^2 \left(\frac{1}{E_0^2}-\frac{1}{(\hbar \omega)^2}\right) \right), \hspace{1.5cm} E_f\leq \hbar\omega\leq E_0
\end{align}
\end{subequations}

\noindent
where $C = 16 Z^2 r_e^2 \alpha n N_e L_0/3$.   

\subsection*{ICS model calculations}

An analytic expression for the energy radiated per unit frequency per unit solid angle by a single electron with Lorentz factor $\gamma_0$, oscillating in a linearly polarized plane wave laser pulse with strength parameter $a_0$ and central frequency $\omega_0$ is derived by Esarey \textit{et al.} (1993):

\begin{equation}\label{ICS big equation}
    \frac{d^2I}{d(\hbar\omega) d\Omega} = \sum_{n=1}^{\infty} \frac{e^2 N_0^2}{16 \pi \hbar \epsilon_0 c}\left(\frac{k}{k_0}\right)^2 Res(k,nk_0) \times F_n
\end{equation}

\noindent
Here, $N_0$ is the number of laser periods, $n$ the harmonic number, $k_0$ the scattering wave number of the laser and $k$ is the emitted wave number of the radiation. The radiation for a given harmonic is peaked at a resonant frequency defined by the resonant function, $Res(k,nk_0)$:

\begin{equation}\label{resonant function}
    Res(k,nk_0) = \left[\frac{\sin{\bar{k} L/2}}{\bar{k} L/2}\right]^2
\end{equation}

\noindent
where $\bar{k} = nk_0 - k(1+a_0^2/2 +\gamma_0^2\theta^2)/4\gamma_0^2$, $L$ is the length of the laser pulse $c\tau$ and $\theta$ is the observation angle eith respect to the axis. This results in a resonant frequency of

\begin{equation}\label{resonant frequency}
    \omega_n = \frac{4\gamma_0^2 \omega_0 n}{1+a_0^2/2 + \gamma_0^2\theta^2}
\end{equation}

\noindent
The relative width of the resonant function is $\Delta\omega / \omega_n = 1/nN_0$ and for a 800 nm, 30 fs pulse $N_0=c\tau/\lambda_0\approx 10$, contributing about 10\% of the relative energy spread for the fundamental, $n=1$. 

The $F_n$ component of equation\,\eqref{ICS big equation} is a harmonic amplitude function which gives the relative weight of a harmonic as a function of $a_0$, $\theta$ and $\gamma_0$ \cite{Esarey1993NonlinearPlasmas}. For $a_0<<1$ the only significant component is the fundamental, but as early as $a_0\sim0.3$ the second harmonic can play a significant role when integrating over observation angles. To account for electron energy spread and divergence, we integrate the single electron energy over the electron's phase space $N_e f(\gamma,\theta_e)$, where $\theta_e$ is the angle the electron travels with respect to the axis and $f(\gamma,\theta_e)$ is normalized to the total electron number $N_e$:

\begin{equation}\label{ICS integrals}
    \frac{dI_{tot}}{d(\hbar\omega)} = 2\pi \int_0^{\theta_{max}} \sin\theta d\theta \int d\theta_e \int  f(\gamma,\theta_e) \frac{d^2I}{d(\hbar\omega) d\Omega}(\theta-\theta_e,\gamma) d\gamma.
\end{equation}

\begin{figure}[!ht]
    \centering
    \includegraphics[width=.95\textwidth]{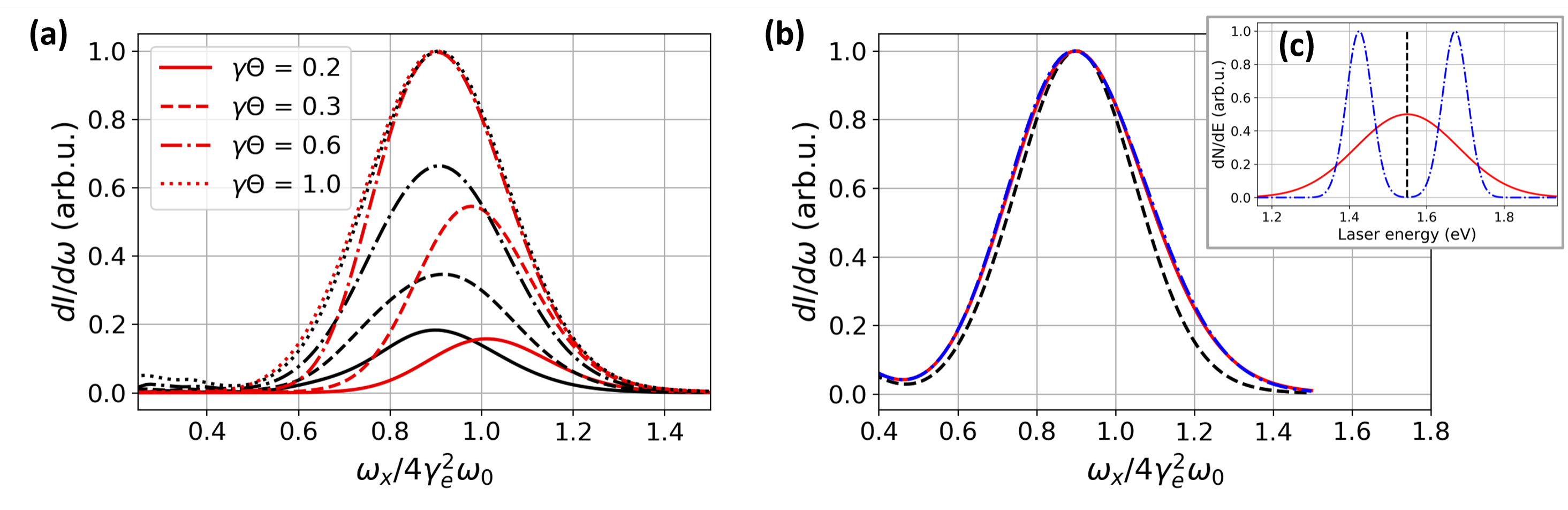}
    \caption{\label{fig:ICS supp plot 1}(a) The calculated ICS spectra for $a_0=0.01$ and laser frequency $\omega_0$ from an electron bunch with 6.5\% energy spread and no divergence (red curves) and from a bunch with divergence $\sigma_e=1.7$ mrad (black curves). Each curve is the result of integrating equation\,\eqref{ICS integrals} up to $\gamma_0\theta_{max}$ of 0.2 (solid), 0.3 (dashed), 0.6 (dot-dashed) and 1.0 (dotted). (b) The calculated ICS spectrum from non-divergent electrons integrated up to $\gamma_0\theta_{max}=1$ from a single laser frequency of $\hbar\omega_0=$1.55 eV (black dashed), a Gaussian laser spectrum with 8.5\% energy spread (red solid) and from a laser spectrum with two peaks separated by 9\% (blue dot-dashed). The laser spectra that correspond to the calculations in (b) are shown in (c).}
\end{figure}

The integration over $\theta_e$ and $\gamma$ can be limited to the extent of the beam, and the maximum observation angle, $\theta_{max}$ is chosen as the acceptance angle for the stack measurements. Electrons that travel at an angle $\theta \neq 0$ with respect to the axis will still emit radiation on axis that is redshifted according to equation\,\eqref{resonant frequency}. The inclusion of $\theta_e$ simply acts to average the angular distribution at \emph{any} observation angle. This effect is illustrated in Fig.\,\ref{fig:ICS supp plot 1}(a). The total energy radiated from an electron bunch with a peak energy of 250 MeV ($\gamma_0=490$) and $\sigma_\gamma/\gamma_0 = 0.065$ is calculated assuming a non-divergent electron bunch (red curves) and an electron bunch with 2-D gaussian profile and $\sigma_e = 1.7$ mrad (black curves). The electron divergence and energy spread were chosen to closely match the measured bunch parameters. The distribution is integrated up to a normalized angle $\gamma\theta$ of 0.2 (solid), 0.3 (dashed), 0.6 (dot-dashed) and 1 (dotted) to clearly illustrate the effect of the electron divergence. The non-divergent electron case (red curves) illustrate the redshifting of the spectrum that occurs at off-axis observation angles, starting at $\omega_x \approx 4\gamma_0^2 \omega_0$ and reducing by about 10\% after integrating over $\gamma\theta=1$. The divergent electron case (black curves) maintain the same peak location and width for all final integration values and only increases in amplitude as more electrons are included in the summation. The resulting shape is the same as that reached from a non-divergent electron bunch at $\gamma\theta=1$. By dropping the additional integral over $\theta_e$ the final spectrum remains valid when for a divergent electron bunch as long as the integration over observation angles extends to $\gamma \theta = \gamma \theta_e$.

\noindent
After dropping the integration over $\theta_e$ we can write the total energy radiated from an electron bunch per unit $\hbar \omega$ as

\begin{equation}\label{dIdhw}
    \begin{split}
       \left\langle \frac{ d I_{tot}}{d(\hbar\omega)} \right\rangle_\phi & = 2 \pi \alpha N_e N_0^2 a_0^2\int_0^{\theta_{max}} \sin\theta d\theta \int d\gamma f(\gamma) \gamma^2 \left( \frac{1 + \gamma^4 \theta^4 }{\left( 1 + \gamma^2 \theta^2 \right)^4}\right) Res(k,n k_0)
    \end{split}
\end{equation}

\noindent
where we have assumed $a_0 \ll 1$, $\theta \ll 1$ and averaged over the azimuthal angle $\phi$.

Another factor that slows down computation is the inclusion of a realistic scattering laser spectrum. To implement this feature, an integral over the laser spectrum needs to be added. Additionally, the central laser wavelength is expected to redshift as the laser pulse depletes and energy is transferred to the wake. The depletion length for these accelerators is \cite{Lu2007GeneratingRegime}

\begin{equation}\label{Ldep}
    L_{pd} \simeq \frac{\omega_0^2}{\omega_p^2} c\tau = \frac{n_c}{n_e} c\tau
\end{equation}

\noindent
where $n_c$ is the critical density and $c\tau$ is the spatial extent of the laser pulse. The redshift can be estimated from the depletion length as

\begin{equation}\label{Lu eq}
    \frac{\Delta\omega}{\omega_0} \sim \frac{L_{acc}}{L_{pd}}
\end{equation}

For conditions in the experiment, $n_c=1.7\times10^{21} \mathrm{cm^{-3}}$ and $n_e =  4\times10^{18} \mathrm{cm^{-3}}$ and $n_e =6\times10^{18} \mathrm{cm^{-3}}$ for shot 1 and 2 in the ICS section of the main text, respectively. The acceleration length in the STII regime is typically on the order of 1 mm and the depletion length is between 4 and 3 mm respectively resulting in a shift of 25\% and 40\%. Moreover, the nominal energy spread of the laser is $\sigma_L/E_L$ is $\sim0.03$ and the broadening from the LWFA is expected to be primarily caused by the laser depletion and redshift. Structures within the laser spectrum do not impart the same features on the resulting ICS spectrum as long as the spread remains less than the relative broadening caused by the electron divergence and energy spread as well as the angular integration. Fig.\,\ref{fig:ICS supp plot 1}(b) shows the calculated ICS spectra integrated over $\theta\gamma = 1$ from an electron bunch ($\sigma_\gamma/\gamma_0 = 0.065$) scattering from the three different laser spectra shown in \ref{fig:ICS supp plot 1}(c). The nominal laser assumes a single laser frequency $E_{L0} = \hbar\omega_0 = 1.55$ eV (black dotted) and compares it with a Gaussian spectrum centered at $E_{L0}=1.55$ eV and 8.5\% spread (red solid) and a spectrum with two peaks placed at $E_{L0}\pm0.9E_{L0}$ and 2\% spread (blue dot-dashed). The cases assuming $\sim9\%$ laser spread result in the same calculated ICS spectrum regardless of the relative shape of the spectrum. Furthermore, they add only $\sim2\%$ to the total energy spread making the central frequency the primary feature of the laser pulse to include in the calculation. Thus, for linear ICS where $a_0 \ll 1$, the energy spectrum can be approximated as a Gaussian function with mean photon energy $E_x$ and variance $\sigma_{E_x}$. 

\bibliography{references_static}